\documentclass[a4paper, 12pt]{article}

\usepackage{times}

\usepackage{romannum}
\usepackage{amsmath}
\usepackage{titling}
\usepackage{blindtext}
\usepackage{amsfonts}
\usepackage{amsthm}
\usepackage{amssymb}
\usepackage[round]{natbib}

\usepackage[linesnumbered, ruled, vlined]{algorithm2e}
\usepackage[utf8]{inputenc}
\usepackage[english]{babel}
\usepackage{enumitem}
\usepackage{accents}
\usepackage{url}
\usepackage{mathtools}  
\usepackage{mathrsfs}   
\usepackage[margin=1in]{geometry}

\usepackage{graphicx}
\graphicspath{ {./figures/} }

\newtheorem{theorem}{Theorem}

\theoremstyle{definition}
\newtheorem{definition}{Definition}

\newcommand{\keyword}{\textit{Keywords: }}
\newcommand{\bhtheta}{\widehat{\boldsymbol{\theta}}_n}
\newcommand{\btheta}{\boldsymbol{\theta}}
\newcommand{\cond}{\xrightarrow{d}}
\newcommand{\utilde}{\underaccent{\tilde}}
\DeclareMathOperator{\sgn}{sgn}

\newenvironment{sciabstract}{%
\begin{quote} }
{\end{quote}}

\newcounter{lastnote}

\title{Predicting the Number of Future Events}

\author{Qinglong Tian, Fanqi Meng, Daniel J. Nordman, William Q. Meeker\\
\\
\normalsize{Department of Statistics, Iowa State University, Ames, IA 50011}
}

\date{\today}

\begin{document} 

\pagenumbering{arabic}
\baselineskip24pt


\maketitle 


\begin{sciabstract}
This paper describes prediction methods for
the number of future events from a population of units associated with
an on-going time-to-event process. Examples include the prediction
of warranty returns and the prediction of the number of future
product failures that could cause serious threats to property or
life. Important decisions such as whether a product recall should be
mandated are often based on such predictions.
Data, generally right-censored (and sometimes left truncated and
right-censored), are used to estimate the parameters of a time-to-event
distribution. This distribution can then be used to
predict the number of events over future periods of
time. Such predictions are sometimes called within-sample
predictions and differ from other
prediction problems considered in most of the prediction literature.
This paper shows that the plug-in (also known as estimative or naive)
prediction method is not asymptotically correct (i.e., for large
amounts of data, the coverage probability always fails to converge to
the nominal confidence level). However, a commonly used
prediction calibration method is shown to be asymptotically correct
for within-sample predictions, and two alternative predictive-distribution-based
methods that perform better than the calibration method are presented and justified.

\end{sciabstract}

\keyword{Binomial predictand; Bootstrap; Calibration; Censored data; Predictive distribution}

\newpage


\section{Introduction}
\label{m_exampls}
There are many applications where it is necessary to predict the
number of future events from a population of units associated with
an on-going time-to-event process. Such applications also require
a prediction interval to quantify statistical prediction uncertainty
arising from the combination of process variability and parameter
uncertainty. Some motivating applications are given below.

\noindent\textbf{Product-A Data}:
This example is from \citet{elawqm1999}, where, during a particular month,
$n$=10000 units of Product-A were put into service. Over the next
48 months, 80 failures occurred and the failure times were recorded.
A prediction interval on the number of failures among the remaining
9920 units during the next 12 months was requested by the management.

\noindent\textbf{Heat Exchanger Tube Data}:
This example is based on data described in \citet{nelson2000}.
Nuclear power plants have steam generators that contain many
stainless steel heat-exchanger tubes. Cracks initiate and grow
in the tubes due to a stress-corrosion mechanism over time. Periodic
inspections of the tubes are used to detect cracks. Consider a fleet
of steam generators having a total of $n$=20,000 tubes. One crack was
detected after the first year of operation, which was followed by
another crack during the second year and six more cracks during the
third year. The data are interval-censored as the exact initiation times
are unknown. A prediction interval was needed for the number of tubes that would
crack from the end of the third year to the end of the tenth year.

\noindent\textbf{Bearing-Cage Data}:
The bearing-cage failure-time data are from \citet{weibullhandbook}
and are provided in the online supplementary material.
Groups of aircraft engines employing this bearing cage were put into
service over time (staggered entry).  At the data freeze date, 6
bearing-cage failures had occurred while the remaining 1697 units
with various service times were still in service (multiple
right-censored data).  To assure that a sufficient number of spare
parts would be available to repair the aircraft engine in a timely
manner, management requested a prediction interval for
the number of bearing-cages that would fail in the next year,
assuming 300 hours of service for each aircraft.
\medskip

The purpose of this paper is to show how to construct prediction
intervals for the number of future events from an on-going time-to-event
process, investigate the properties of different prediction methods,
and give recommendations on which methods to use.

This paper is organized as follows. Section~\ref{background} provides concepts and
background for prediction inference. Section~\ref{single_cohort_within_sample_pred} describes the single-cohort
within-sample prediction problem. Section~\ref{plugin_not_regular} defines how the within-sample
prediction is irregular and demonstrates that the plug-in method fails
to provide an asymptotically correct prediction interval. Section~\ref{calibration}
describes the calibration method for prediction intervals and establishes
its asymptotic correctness. Section~\ref{pred:method} presents two other prediction
interval methods based on predictive distributions. The first one is
a general method using parametric bootstrap samples, while the second
method is inspired by generalized pivotal quantities and applies to a
log-location-scale family of distributions. Section~\ref{sec:multiple-cohort} extends the
single-cohort within-sample prediction to the multiple-cohort problem.
Section~\ref{simu:study} compares different prediction methods, through simulation,
while Section~\ref{sec:applications} applies the prediction methods to the motivating examples.
Section~\ref{choice-of-dist} discusses the choice of distribution for the time-to-event
process and addresses the issue of distribution misspecification.
Section~\ref{sec:conclusion} gives recommendations and describes potential areas for future
research.

\section{Background}
\label{background}
In a general prediction problem, denote the observable data by
$\boldsymbol{D}_n$ and the future random variable by $Y_n\equiv Y$;
while generic for now, later this paper will focus on the
within-sample prediction where $Y$ is a count.
The conditional cdf for $Y$ given $\boldsymbol{D}_n$ is denoted by $G_n(\cdot|\boldsymbol{D}_n;
\boldsymbol{\theta})\equiv G(\cdot|\boldsymbol{D}_n;
\boldsymbol{\theta})$, where $\boldsymbol{\theta}$ is a vector of parameters.
The goal is to make inference for $Y$ through a prediction interval,
as a useful tool for quantifying uncertainty in prediction.

\subsection{Prediction Intervals}
\label{predinterval}
When parameters in $\btheta$ are known, the one-sided upper $100(1-\alpha/2)\%$
prediction bound $\tilde{Y}_{1-\alpha/2}$ is defined as the $100(1-\alpha/2)\%$
quantile of the conditional cdf for $Y$,
which is
\begin{equation}
\tilde{Y}_{1-\alpha/2}=\inf\{y\in\mathbb{R}:G(y|\boldsymbol{D}_n;\btheta)=\Pr(Y\leq y\vert \boldsymbol{D}_n, \btheta)\geq1-\alpha/2\},
\label{upperbound::true}
\end{equation}
and the one-sided lower $100(1-\alpha/2)\%$ prediction bound may be defined as
\begin{equation}
\utilde{Y}_{1-\alpha/2}=\sup\{y\in\mathbb{R}:\Pr(Y\geq y\vert \boldsymbol{D}_n, \btheta)\geq1-\alpha/2\},
\label{lowerbound::true}
\end{equation}
where this modification of the usual $\alpha/2$ quantile of $Y$
ensures that $\Pr(Y\geq\utilde{Y}_{1-\alpha/2}|\boldsymbol{D}_n, \btheta)$
is at least $100(1-\alpha/2)\%$ when $Y$ is a discrete random variable.
We may obtain an equal-tail $100(1-\alpha)\%$ prediction interval
(approximate when $Y$ is a discrete random variable) by combining
these two prediction bounds.

In most applications, equal-tail prediction intervals are preferred
over unequal ones, even though it is sometimes possible to find a
narrower prediction interval with unequal tail probabilities.
This is because the equal-tail prediction interval can be naturally
decomposed into a practical one-sided upper prediction bound and a lower
prediction bound where the separate consideration of one-sided bounds is needed
when the cost of being outside the prediction bound is much higher
on one side than the other.

When the parameters in $\btheta$ are unknown,
an estimation of $\btheta$ from the observed data $\boldsymbol{D}_n$ is required.
The plug-in method, also known as the naive or estimative method
(cf.~Section~\ref{literature_review}), is to replace $\btheta$
with a consistent estimator $\bhtheta$ in the prediction bounds (\ref{upperbound::true}) and (\ref{lowerbound::true}).
The $100(1-\alpha)\%$ plug-in upper prediction bound is then
$\tilde{Y}_{1-\alpha}^{PL}=\inf\{y\in\mathbb{R}:G(y|\boldsymbol{D}_n;\bhtheta)\geq1-\alpha\}$
while the $100(1-\alpha)\%$ plug-in lower prediction bound is
$\utilde{Y}_{1-\alpha/2}^{PL}=\sup\{y\in\mathbb{R}:\Pr(Y\geq y\vert \boldsymbol{D}_n, \bhtheta)\geq1-\alpha\}$.

\subsection{Coverage Probability}
\label{coverageprob}
Besides the plug-in method, other methods for computing
prediction bounds or intervals are available.
Let $\mathrm{PI}(1-\alpha)$ generically denote a
prediction interval (or bound) of a nominal coverage level
$100(1-\alpha)\%$, where researchers would like the probability of $Y$
falling within the interval to be (or close to) $1-\alpha$
(i.e., $\Pr[Y\in\mathrm{PI}(1-\alpha)]=1-\alpha$).

To be clear, there are two possible types of coverage probability:
conditional coverage probability and unconditional (overall) coverage probability.
The conditional coverage probability of a particular
$\mathrm{PI(1-\alpha)}$ method is defined as
\begin{align*}
\mathrm{CP}[\mathrm{PI}(1-\alpha)| \boldsymbol{D}_n; \btheta]=\Pr[Y\in\mathrm{PI}(1-\alpha)| \boldsymbol{D}_n; \btheta],
\end{align*}
where $\Pr(\cdot|\boldsymbol{D}_n; \btheta)$ denotes the
conditional probability of $Y$ given the observable data
$\boldsymbol{D}_n$. The conditional coverage probability
$\mathrm{CP}[\mathrm{PI}(1-\alpha)|\boldsymbol{D}_n; \btheta]$
is a random variable because it is a function of the data
$\boldsymbol{D}_n$.
The unconditional coverage probability of a prediction interval
method can be obtained by taking an expectation with respect to
the data $\boldsymbol{D}_n$ and it is defined as
\begin{align*}
\mathrm{CP}[\mathrm{PI}(1-\alpha); \btheta]=\boldsymbol{\mathrm{E}}\left\{\Pr[Y\in\mathrm{PI}(1-\alpha)| \boldsymbol{D}_n; \btheta]\right\}.
\end{align*}
The unconditional coverage probability is a fixed property of a
prediction method and, as such, can be most readily studied and used to compare
alternative prediction interval methods.
We focus on unconditional coverage probability in this paper and use the
term coverage probability to refer to the unconditional probability,
unless stated otherwise.

We say a prediction method is exact if
$\mathrm{CP}[\mathrm{PI}(1-\alpha); \btheta]=1-\alpha$ holds.
If $\mathrm{CP}[\mathrm{PI}(1-\alpha); \btheta]$ converges to
$1-\alpha$ as the sample size $n$ increases,
we say the corresponding prediction method is asymptotically correct.
When $Y$ is a discrete random variable, however, asymptotic
correctness and exactness may not generally hold or be possible
for a prediction interval method, due to the discreteness in the distribution of $Y$.

\subsection{Related Literature}
\label{literature_review}
Extensive research exists regarding some methods for computing
prediction intervals. While the plug-in method has been criticized
for ignoring the uncertainty in $\bhtheta$, this method is often
widely viewed as being asymptotically correct (related to
``regular predictions'' described in Section~\ref{regular_prediction}).
For example, \citet{cox1975}, \citet{beran1990}, and \citet{hall1999}
showed that the coverage probability of the plug-in method has an
accuracy of $O(n^{-1})$ for a continuous predictand under
certain conditions. In Section~\ref{plugin_not_regular} we show,
however, that the plug-in method is not asymptotically correct
in the context of within-sample prediction.

Section~\ref{calibration} presents a calibration method for
within-sample prediction intervals. \citet{cox1975} originally
proposed the calibration idea to improve on the plug-in method
and also provided analytical forms for prediction intervals
based on general asymptotic expansions. \citet{atwood1984} 
used a similar method. \citet{beran1990} employed bootstrap
in the calibration method, avoiding the complicated analytical
expressions. \citet{elawqm1999} described similar methods for
constructing prediction intervals for failure times and the number
of future failures, based on censored life data.

This paper does not specifically address Bayesian prediction methods,
but the classic idea of a Bayesian predictive
distribution can be extended to non-Bayesian methods and
two such methods are considered in Section~\ref{pred:method}. Several
authors have considered similar notions of a non-Bayesian
predictive distribution (e.g., \citet{aitchison1975},
\citet{davison1986}, \citet{barncox1996}). \citet{lawless2005}
demonstrated a relationship between predictive distributions
and (approximate) pivotal-based prediction intervals, including
the calibration method described in \citet{beran1990}.
\citet{fonseca2012} further elaborated on the relationship
between predictive distributions and the calibration method.
\citet{shen_liu_xie_2018} proposed a general framework to
construct a predictive distribution by replacing the
posterior distribution in the definition of a Bayesian
predictive distribution with a confidence distribution.


\section{Single Cohort Within-Sample Prediction}
\label{single_cohort_within_sample_pred}
\subsection{Within-Sample Prediction and New Sample Prediction}
The term ``within-sample'' prediction has been used to distinguish from
the more widely known ``new sample'' prediction. In new-sample prediction,
past data are used, for example, to compute a prediction interval for the lifetime of a
single unit from a new and completely independent sample.
For within-sample prediction, however, the sample has not changed;
the future random variable that researchers wish to predict (i.e., a count)
relates to the same sample that provided the original (censored) data.
\subsection{Single-Cohort Within-Sample Prediction and Plug-in Method}
\label{withinsample}
Let $({T}_1,...,{T}_n)$ be an unordered random sample from
a parametric distribution $F(t;\btheta)$ having support on the positive real line
and $\btheta\in\mathbb{R}^q$.
Under Type~\Romannum{1} censoring at $t_c>0$,
the available data may then be expressed by $D_i=(\delta_i, T_i^{obs}),i=1,...,n$,
where $\delta_i=\mathrm{I}(T_i\leq t_c)$ is a variable indicating
whether $T_i$ is observed before the censoring time $t_c$,
so that the actual observed variables are given as
$T_i^{obs}=T_i\delta_i+t_c(1-\delta_i)$.
The observed number of events (uncensored units) in the sample will
be denoted by $r_n=\sum_{i=1}^{n}\mathrm{I}(T_i\leq t_c)$.
For a future time $t_w>t_c$, let $Y_n=\sum_{i=1}^{n}\mathrm{I}(T_i\in(t_c, t_w])$
denote the (future) number of values from $T_1,...,T_n$, that
occur in the interval $(t_c, t_w]$. The conditional distribution
of $Y_n$ is then $\mathrm{binomial}(n-r_n, p)$ given the observed
data $\boldsymbol{D}_n=(D_1,...,D_n)$, where $p$ is the conditional
probability that $T_i\in(t_c, t_w]$ given that $T_i>t_c$.
As a function of $\btheta$, we may define $p$ by
\begin{equation}
p\equiv\pi(\btheta)=\frac{F(t_w;\btheta)-F(t_c;\btheta)}{1-F(t_c;\btheta)}.
\label{piequa}
\end{equation}
The goal is to construct a prediction interval for $Y_n$ based on
the observed data $\boldsymbol{D}_n=(D_1,...,D_n)$ when $\btheta$ is unknown.
This is referred to as single-cohort within-sample prediction because
all the units enter the system at the same time and are homogeneous;
and both the data $\boldsymbol{D}_n$ and the predictand $Y_n$ are
functions of the uncensored random sample $(T_1,...,T_n)$.

Let $\bhtheta$ denote an estimator of $\btheta$ based on $\boldsymbol{D}_n$,
then a plug-in estimator $\widehat{p}_n=\pi(\bhtheta)$ of the
conditional probability $p$ follows from (\ref{piequa}).
Analogous to the bounds in Section~2.1, a $100(1-\alpha)\%$ plug-in lower prediction bound is defined as
\begin{align*}
\utilde{Y}^{PL}_{n, 1-\alpha}&=\sup\{y\in\{0\}\cup\mathbb{Z}^{+}; \mathrm{pbinom}(y-1, n-r_n, \widehat{p}_{n})\leq\alpha\}
\\&=
\begin{cases}
\mathrm{qbinom}(\alpha, n-r_n, \widehat p_n), &\text{if } \mathrm{pbinom}(\mathrm{qbinom}(\alpha, n-r_n, \widehat p_n), n-r_n, \widehat p_n)>\alpha.\\
\mathrm{qbinom}(\alpha, n-r_n, \widehat p_n)+1, &\text{if } \mathrm{pbinom}(\mathrm{qbinom}(\alpha, n-r_n, \widehat p_n), n-r_n, \widehat p_n)=\alpha.\\
\end{cases}
\end{align*}
where $\mathrm{pbinom}$ and $\mathrm{qbinom}$ are, respectively,
the binomial cdf and quantile function.
Similarly, the $100(1-\alpha)\%$ plug-in upper prediction bound for $Y_n$ is defined as
\begin{align*}
\tilde{Y}^{PL}_{n, 1-\alpha}&=\inf\{y\in\{0\}\cup\mathbb{Z}^{+};
\mathrm{pbinom}(y, n-r_n, \widehat{p}_{n})\geq1-\alpha\}=\mathrm{qbinom}(1-\alpha, n-r_n, \widehat p_n).
\end{align*}

Section~\ref{coverageprob} mentioned that asymptotically
correct coverage may not generally be possible for prediction
intervals involving a discrete predictand. However, for within-sample
prediction here, prediction interval methods can be sensibly examined
for properties of asymptotic correctness,
which we consider in the following section.
This is because discreteness in the (conditionally) binomial predictand
$Y_n$ essentially disappears in large sample sizes $n$, due to
normal approximations.
\section{The Irregularity of the Within-Sample Prediction}
\label{plugin_not_regular}
\subsection{A Regular Prediction Problem}
\label{regular_prediction}
Under the general prediction framework described in Section~\ref{background},
the conditional cdf $G_n(\cdot| \boldsymbol{D}_n;\btheta)$ of a predictand
$Y_n$ given the observed data $\boldsymbol{D}_n$ is often estimated by the plug-in
method as $G_n(\cdot| \boldsymbol{D}_n;\bhtheta)$ (also known as a predictive distribution),
where $\bhtheta$ is a consistent estimator of $\btheta$ based on $\boldsymbol{D}_n$.
To frame much of the literature related to the plug-in method (Section~\ref{literature_review}),
we may define the prediction problem most often commonly related to the plug-in method
as ``regular'' according to the following definition.
\begin{definition}
In the notation of Section~\ref{background}, a prediction problem is called regular if
\begin{equation*}
\sup_{y\in\mathbb{R}}|G_n(y|\boldsymbol{D}_n; \btheta)-G_n(y|\boldsymbol{D}_n; \bhtheta)|\xrightarrow{p}0
\end{equation*}
holds as $n\to\infty$ for any consistent estimator $\bhtheta$ of $\btheta$ (i.e., $\bhtheta\xrightarrow{p} \btheta$).
\label{def1}
\end{definition}
Unlike coverage probability (where exactness may again not be possible for discrete predictands),
the above definition reflects the underlying sense of how the plug-in method for prediction
intervals is often asymptotically valid for both discrete and continuous predictands.
By the nature of many prediction problems (e.g., new sample prediction),
the conditional form of cdf $G_n$ may also not necessarily vary with $n$
(e.g., $G_n( \cdot |\boldsymbol{D}_n;\btheta )= G( \cdot; \btheta)$).
Hence, in a regular prediction problem, the plug-in predictive distribution (estimated cdf) asymptotically captures the true conditional cdf of the predictand, so that differences are expected
to vanish between quantiles of the true predictand $Y_n$ and the associated plug-in prediction bounds.
Further, when the predictand has a continuous and asymptotically tight
conditional distribution (with probability 1),
such as when the conditional cdf $G_n(\cdot| D_n; \btheta) = G(\cdot;\btheta)$
of the predictand does not vary with $n$,
then the plug-in method will be asymptotically correct.

\subsection{Failure of the Plug-in Method}

This section shows that the within-sample prediction problem described in Section~\ref{single_cohort_within_sample_pred} is not
regular and that the plug-in method is not asymptotically valid for
within-sample prediction.
To avoid redundancy, the presentation of results will focus on the plug-in
upper prediction bound; the lower bound is analogous by Remark~1 below.
In the context of within-sample prediction (cf.~Section~\ref{withinsample}),
recall that the $100(1-\alpha)\%$ plug-in upper prediction bound
for the future count $Y_n \equiv \sum_{i=1}^n \mathrm{I}(T_i \in (t_c,t_w])$ is defined as 
\begin{align*}
\tilde{Y}^{PL}_{n, 1-\alpha}=\inf\{y\in\mathbb{Z}; \mathrm{pbinom}(y, n-r_n, \widehat{p}_{n})\geq1-\alpha\}.
\end{align*}
The following theorem shows that the coverage probability of
$\tilde{Y}^{PI}_{n, 1-\alpha}$ will not correctly converge to
$1-\alpha$ as $n$ increases.
\begin{theorem}
Let $T_{1}, ..., T_{n}$ denote a random sample from a parametric distribution
with cdf $F(\cdot; \boldsymbol{\theta}_{0})$ (at the true value of $\btheta=\boldsymbol{\theta}_{0}\in\mathbb{R}^q$),
which is observed under Type~I censoring at $t_c>0$.
Suppose also that $F(t_c;\btheta_0) <1$, $p_0 = \pi(\btheta_0)\in(0, 1)$ in (\ref{piequa}), $F(t_c;\btheta)$ is continuous at $\btheta_0$, and that the conditional probability (parametric function) $p\equiv\pi(\btheta)$ is continuously differentiable in a neighborhood of $\btheta_0$ with non-zero gradient $\nabla_0\equiv\partial \pi(\btheta)/\partial \btheta|_{\btheta=\btheta_{0}}$.
Based on the censored sample, suppose $\bhtheta$ is an
estimator of $\btheta$ satisfying $\sqrt{n}(\bhtheta-\btheta_{0})\cond \mathrm{MVN}(\boldsymbol{0}, \boldsymbol{V}_0)$,
as $n\rightarrow\infty$, for a multivariate normal distribution with mean vector $\boldsymbol{0}$ and positive definite variance matrix $\boldsymbol{V}_{0}$.
Then,
\begin{enumerate}
\item The within-sample prediction of $Y_n = \sum_{i=1}^n \mathrm{I}(t_c < T_i \leq t_w)$ fails to be a regular prediction problem:
denoting $G_n(y|\boldsymbol{D}_n,\btheta_0)=\text{pbinom}(y,n-r_n,p_0)$ as the conditional cdf of $Y_n$ and $G_n(y|\boldsymbol{D}_n,\bhtheta)=\text{pbinom}(y,n-r_n,\widehat{p}_n)$ as its plug-in estimator, then
\[
\sup_{y \in \mathbb{R}} \left| G_n(y|\boldsymbol{D}_n,\btheta_0) -  G_n(y|\boldsymbol{D}_n,\bhtheta)\right|
\xrightarrow{d} 1 -2\Phi_{\mathrm{nor}}(\sqrt{v_1}|Z_1|/2),
\]
where $Z_1$ is a standard normal variable with cdf $\Phi_{\mathrm{nor}}(z)=\int_{-\infty}^{z} 1/\sqrt{2 \pi} e^{-u^{2} / 2}d u$, $z\in\mathbb{R}$, and
$$
v_{1}\equiv\frac{[1-F(t_{c}; \btheta_{0})]}{p_{0}(1-p_{0})}\nabla_{0}^{t}\boldsymbol{V}_{0}\nabla_0\in(0, \infty).
$$
\item The plug-in upper prediction bound $\tilde{Y}^{PL}_{n, 1-\alpha}$ generally fails to have asymptotically correct coverage:
\begin{align*}
\lim_{n\rightarrow\infty}\Pr(Y_{n}\leq \tilde{Y}^{PL}_{n, 1-\alpha})=\Lambda_{1-\alpha}(v_1)\in(0,1)
\quad
\text{such that}
\\
\sgn\left[\Lambda_{1-\alpha}(v_1)-(1-\alpha)\right]=
\begin{cases}
1&\quad\mbox{if $\alpha \in(1/2,1)$}\\
0&\quad\mbox{if $\alpha=1/2$}\\
-1&\quad\mbox{if $\alpha\in(0,1/2)$},
\end{cases}
\end{align*}

where $\sgn(\cdot)$ is the sign function and $\Lambda_{1-\alpha}(v_1) \equiv \int_{-\infty}^{\infty}\Phi_{\mathrm{nor}}\left[\Phi_{\mathrm{nor}}^{-1}(1-\alpha)+z \sqrt{v_{1}}\right]d \Phi_{\mathrm{nor}}(z)$.
Furthermore, $\Lambda_{1-\alpha}(v_1) \in [1/2,1-\alpha)$ is a decreasing function of $v_1>0$
for a given $\alpha \in (0,1/2)$, while $\Lambda_{1-\alpha}(v_1) \in (1-\alpha,1/2]$
is increasing in $v_1>0$ for $\alpha \in (1/2,1)$,
and $\lim_{v_1 \to \infty}\Lambda_{1-\alpha}(v_1)=1/2$ holds for any $\alpha\in(0,1)$.
\end{enumerate}
\label{first_theorem}
\end{theorem}

\noindent\textbf{Remark 1}. The lower plug-in bound $\utilde{Y}_{n,1-\alpha}^{PL}$ behaves similarly with $\lim_{n\to \infty}\Pr(Y_n \geq\utilde{Y}_{n, 1-\alpha}^{PL}) = \lim_{n\to\infty}\Pr(Y_n \leq \tilde{Y}_{n,1-\alpha}^{PL})$ in Theorem 1.\\
\indent The proof of Theorem~\ref{first_theorem} is in the online supplementary material.
This counter-intuitive result reveals that the plug-in method should not
be used to construct prediction intervals in the within-sample prediction
problem, even if the sample size is large.
The first part of Theorem~\ref{first_theorem} entails that plug-in estimation
fails to capture the distribution of the predictand $Y_n$ here, to the
extent that the supremum difference between estimated and true distributions
has a {\em random} limit, rather than converging to zero as in a regular
prediction (cf.~Definition~\ref{def1}). As a consequence, the limiting
coverage probability of the plug-in bound turns out to be ``off'' by an
amount determined by a magnitude of $v_1>0$ in Theorem~\ref{first_theorem} (part 2).
For increasing values of $v_1$, the coverage probability approaches 0.5,
regardless of the nominal coverage level intended.
An intuitive explanation for the failure of plug-in method is that, although $\widehat{p}_{n}$ converges consistently to $p$, the growing number of Bernoulli trials $n-r_n$ in $Y_{n}$ offsets the improvements that larger samples may offer in estimation by $\widehat p_n$.
In other words, when standardizing the true $1-\alpha$ quantile,
say $Y_{n,1-\alpha}$, of the (conditionally binomial) predictand $Y_n$,
one obtains a standard normal quantile
$(Y_{n,1-\alpha} -p)/\sqrt{n-r_n}\approx \Phi_{\mathrm{nor}}^{-1}(1-\alpha)$ by normal approximation;
however, the same standardization applied to the plug-in bound $\tilde{Y}_{n,1-\alpha}^{PL}$ gives $(\tilde{Y}_{n,1-\alpha}^{PL} -p)/\sqrt{n-r_n}\approx \Phi_{\mathrm{nor}}^{-1}(1-\alpha) + \sqrt{n-r_n} (\widehat{p}_n-p)$, which differs by a substantial and random amount $\sqrt{n-r_n} (\widehat{p}_n-p)$ (having a normal limit itself).
Hence, validity of the plug-in method for within-sample prediction would
require an estimator $\widehat p_n$ such that $\widehat p_n=p+o_p(n^{-1/2})$,
which demands more than what is available from standard $\sqrt{n}$-consistency.

\section{Prediction Intervals Based on Calibration}
\label{calibration}
\subsection{Calibrating Plug-in Prediction Bounds}
\citet{cox1975} suggested an approximation for improving the plug-in method,
which will be described next.
Considering the general prediction problem (cf.~Section~\ref{predinterval}), suppose a future random variable $Y \equiv Y_n$ has a conditional cdf $G_n(\cdot|\boldsymbol{D}_n;\btheta) \equiv G(\cdot|\boldsymbol{D}_n; \btheta)$ given random sample $\boldsymbol{D}_n$ and $\bhtheta$ is a consistent estimator of $\btheta$ from $\boldsymbol{D}_n$.
The coverage probability of the $100(1-\alpha)$\% plug-in upper prediction bound is denoted by $\Pr\left[G(Y|\boldsymbol{D}_n;\bhtheta)\leq 1-\alpha\right]=1-\alpha^\prime$, where $\alpha^\prime$ is generally different from $\alpha$
due to the estimation uncertainty in $\bhtheta$.
The basic idea of the calibration method is to find a level $\alpha^\dagger$ so that the coverage probability $\Pr\left[G(Y|\boldsymbol{D}_n;\bhtheta)\leq1-\alpha^\dagger\right]$ is equal to (or closer to) $1-\alpha$.
The resulting $100(1-\alpha^\dagger)\%$ upper plug-in prediction bound $\tilde{Y}_{n,1-\alpha^\dag}^{PL}$ is called the $100(1-\alpha)\%$ upper calibrated prediction bound.
However, determination of $\alpha^\dagger$ relies on both the
distribution of $Y$ and the sampling distribution of $\bhtheta$,
each of which depends on the unknown parameter $\btheta$.
So instead, $\alpha^\dagger$ is obtained by solving the equation
$
\Pr{}_{\!\!\ast}\left[G(Y^\ast|\boldsymbol{D}_n;\widehat{\boldsymbol{\theta}}^{\ast}_n)\leq1-\alpha^\dagger\right]=1-\alpha
$,
where $\Pr{}_{\!\!\ast}$ denotes bootstrap probability induced
by $Y^\ast\sim G(\cdot|\boldsymbol{D}_n;\bhtheta)$ and by
$\widehat{\boldsymbol{\theta}}^{\ast}_n$ as a bootstrap version of $\bhtheta$;
for example, $\widehat{\boldsymbol{\theta}}^{\ast}_n$ may be based on a bootstrap sample $\boldsymbol{D}_n^*$ found by a parametric bootstrap applied using $\bhtheta$ in the role of the unknown parameter vector $\btheta$.
\citet{beran1990} showed, that under certain conditions, instead of having a coverage error of $O(n^{-1})$, the coverage probability of the calibrated upper prediction bound improves upon the plug-in methods, e.g., $\Pr\left[Y\leq G^{-1}(1-\alpha^\dagger|\boldsymbol{D}_n;\bhtheta)\right]=1-\alpha+O(n^{-2})$.
However, such results for the validity of the calibration method cannot
be applied directly to within-sample prediction because conditions in \citet{beran1990} entail that the prediction problem be regular (cf.~Section~\ref{regular_prediction}), which is not true for the within-sample prediction problem (Theorem~\ref{first_theorem}).
Consequently, the issue of asymptotic correctness for the calibration method needs to be determined for within-sample prediction, as next considered.

\subsection{The Calibration-Bootstrap Method for the Within-Sample Prediction}
The general method in \citet{beran1990} is modified to construct a
calibrated prediction interval for within-sample prediction and
it is called the calibration-bootstrap method in the rest of this paper.
For a bootstrap sample $\boldsymbol{D}^\ast_n$ with $r^\ast_n$
observed events (e.g., from a parametric bootstrap using $\bhtheta$),
we define a random variable set $\left(Y_n^\dagger, n-r_{n}^{\ast},
\widehat{p}_{n}^{\ast}\right)$ where $\widehat{p}_{n}^{\ast}=\pi(
\widehat{\boldsymbol{\theta}}^{\ast}_n)$ is the bootstrap version
of $\widehat{p}_n=\pi(\bhtheta)$ and $Y_n^\dagger\sim
\mathrm{binomial}(n-r_{n}^{\ast}, \widehat{p}_n)$, conditional on $r_n^\ast$.

For the $100(1-\alpha)\%$ lower prediction bound, the calibrated confidence level is
$$
\alpha^{\dagger}_{L}=\sup\{u\in[0, 1]:\Pr{}_{\!\!\ast}\left[\mathrm{pbinom}(Y^\dagger_n, n-r_n^\ast,\widehat p_n^\ast)\leq u\right]\leq\alpha\},
$$
where $\Pr{}_{\!\!\ast}$ is the bootstrap probability induced by
$\boldsymbol{D}^\ast_n$, and then the calibrated $100(1-\alpha)\%$
lower prediction bound is given by $\utilde{Y}_{n,1-\alpha}^C=
\utilde{Y}_{n,1-\alpha^\dagger_L}^{PL}$. For the $100(1-\alpha)\%$
upper prediction bound, the calibrated confidence level is
$$
1-\alpha^\dagger_U = \inf\{u\in[0, 1] :\Pr{}_{\!\!\ast}\!\left[\mathrm{pbinom}(Y_n^\dagger, n-r_{n}^{\ast}, \widehat{p}_{n}^{\ast})\leq u\right]\geq 1-\alpha\},
$$
so that the calibrated $100(1-\alpha)\%$ upper prediction bound is
$\tilde{Y}_{n,1-\alpha}^C=\tilde{Y}_{n,1-\alpha^\dagger_U}^{PL}$.
Here $\utilde{Y}_{n,1-\alpha}^{PL}$ and $\tilde{Y}_{n,1-\alpha}^{PL}$
represent lower and upper plug-in prediction bounds, respectively,
as defined in Section~\ref{withinsample}.


The calibration-bootstrap method involves approximating the distribution of $U=\mathrm{pbinom}(Y_{n}, n-r_n, \widehat{p}_{n})$ with the bootstrap distribution of
$U^\ast=\mathrm{pbinom}(Y_{n}^\dagger, n-r^\ast_n, \widehat{p}_{n}^\ast)$. The bootstrap
distribution of $U^\ast$ is used to calibrate the plug-in method.
The procedure of using the calibration-bootstrap method to construct a prediction interval is described below:
\begin{enumerate}
	\itemsep-0.5em
	\item Compute the maximum likelihood (ML) estimate $\widehat{\boldsymbol{\theta}}_n$ using data $\boldsymbol{D}_n$ and the ML estimate $\widehat{p}_n=\pi(\widehat{\boldsymbol{\theta}}_n)$.
	\item Generate a bootstrap sample $\boldsymbol{D}_n^\ast$ and the number of events is denoted by $r_n^\ast$.
	\item Compute $\widehat{\boldsymbol{\theta}}_n^\ast$ and $\widehat{p}_n^\ast=\pi(\widehat{\boldsymbol{\theta}}_n^\ast)$ using the bootstrap sample $\boldsymbol{D}_n^\ast$.
	\item Generate $y^\ast$ from the distribution $\text{binomial}(n-r^\ast_n, \widehat{p}_n)$ and compute $u^\ast=\mathrm{pbinom}(y^\ast, n-r_n^\ast, \widehat{p}_n^\ast)$.
	\item Repeat step 2-4 for $B$ times and get $B$ realizations of $u^\ast$ as $\{u_1^\ast,\dots,u_B^\ast\}$.
	\item Find the $\alpha$ and $1-\alpha$ quantiles of $\{u_1^\ast,\dots,u_B^\ast\}$, and denote these by $u_\alpha$ and $u_{1-\alpha}$, respectively. The $1-\alpha$ calibrated lower and upper prediction bounds are $\utilde{Y}_{n,1-\alpha}^C=\utilde{Y}_{n,1-u_{\alpha}}^{PL}$
	and $\tilde{Y}^C_{n,1-\alpha}=\tilde{Y}_{n,u_{1-\alpha}}^{PL}$.
\end{enumerate}
The pseudo-code for this algorithm is in the online supplementary material.

Next, the calibration-bootstrap method is shown to be asymptotically correct.
This requires a mild assumption on the bootstrap involved, namely that the parameter estimators $\bhtheta^\ast$ in the bootstrap world provide valid approximations for the sampling distribution of the original data estimators $\sqrt{n}(\bhtheta-\btheta)$, in large samples.
More formally, let $\mathcal{L}_n^* \equiv \mathcal{L}_n^*(\boldsymbol{D}_n)$ denote the probability law of the bootstrap quantity $\sqrt{n}(\bhtheta^\ast-\bhtheta)$ (conditional on the data $\boldsymbol{D}_n$) and let $\mathcal{L}_n$ denote the probability law of $\sqrt{n}(\bhtheta-\btheta)$.
Let $\rho(\mathcal{L}_n, \mathcal{L}_n^\ast)$ denote the distance between these distributions under any metric $\rho(\cdot,\cdot)$ that metricizes the topology of weak convergence (e.g., the Prokhorov Metric).
Also, in the bootstrap re-creation, the probability $\Pr{}_{\!\!\ast}(T_1^* \leq t_c)$ that a bootstrap observation $T_1^\ast$ is observed before the censoring time $t_c$ should be a consistent estimator of $F(t_c;\btheta)$ (e.g., $\Pr{}_{\!\!\ast}(T_1^* \leq t_c) = F(t_c;\bhtheta)$ would hold as a natural estimator under a parametric bootstrap).
\begin{theorem}
Under the conditions of Theorem~\ref{first_theorem}, suppose that $\rho(\mathcal{L}_n^*, \mathcal{L}_n) \stackrel{p}{\rightarrow} 0$ and $\Pr{}_{\!\!\ast}(T_1^* \leq t_c) \stackrel{p}{\rightarrow} F(t_c ; \btheta_0)$ as $n\to \infty$.  Then, the $100(1-\alpha)\%$ calibrated upper and lower prediction bounds, respectively $\tilde{Y}^{C}_{n, 1-\alpha}$ and $\utilde{Y}^{C}_{n,1-\alpha}$ have asymptotically correct coverage, that is
\begin{equation*}
\lim_{n\rightarrow\infty}\Pr(Y_{n}\leq\tilde{Y}^{C}_{n, 1-\alpha}) = 1-\alpha=\lim_{n\rightarrow\infty}\Pr(Y_{n}\geq\utilde{Y}^{C}_{n,1-\alpha}).
\end{equation*}
\label{theocali}
\end{theorem}
The proof is in the online supplementary material.
Theorem~\ref{theocali} and its extension in Section 7
guarantee, for example, that the calibration prediction method employed in \citet{elawqm1999}, \citet{hong2009}, \citet{hong2010}, and \citet{hong2013} to construct the prediction intervals for the cumulative number of events is asymptotically correct.

\section{Prediction Intervals Based on Predictive Distributions}
\label{pred:method}
\subsection{Predictive Distributions}
Under the general prediction setting in Section~\ref{background}, recall that the predictive distribution under the plug-in method, given by $G(\cdot|\boldsymbol{D}_n, \bhtheta)$,  provides an estimator of the conditional cdf $G(\cdot|\boldsymbol{D}_n; \btheta)$,
of the predictand $Y$.
Quantiles of this predictive distribution can be associated with prediction bounds for $Y$.
Generally speaking, any method that leads to a prediction bound for $Y$ can be translated to a predictive distribution by defining the $100(1-\alpha)\%$ upper prediction bound as the $1-\alpha$ quantile of the predictive distribution (and vice versa).
In this section, the strategy is to construct predictive distributions that lead to prediction bound (or interval) methods having asymptotically correct coverage for within-sample prediction.

For this purpose, it is helpful to consider a Bayesian predictive distribution, defined by
\begin{equation}
G_{B}(y |\boldsymbol{D}_n)=\int G(y |\boldsymbol{D}_n; \btheta) \gamma(\btheta |\boldsymbol{D}_n) d \btheta,
\label{bayespred}
\end{equation}
where $\gamma(\btheta |\boldsymbol{D}_n)$ is a joint posterior distribution for $\btheta$.
The $1-\alpha$ quantile of the Bayesian predictive distribution provides the $100(1-\alpha)\%$ upper Bayesian prediction bound.
While this paper does not pursue the Bayesian method, the idea of
the Bayesian predictive distribution can nevertheless
be used by replacing the posterior $\gamma(\btheta |\boldsymbol{D}_n)$
in (\ref{bayespred}) with an alternative distribution over
parameters to similarly define non-Bayesian predictive distributions.
\citet{harris1989} replaced the posterior distribution in (\ref{bayespred}) with the bootstrap distribution of the parameters to construct a predictive distribution
while \citet{wang2012} replaced the posterior distribution with a fiducial distribution.
\citet{shen_liu_xie_2018} proposed a framework for predictive inference by replacing the posterior distribution in (\ref{bayespred}) with a confidence distribution (CD) and provided theoretical results for this CD-based predictive distribution for the case of a scalar parameter.
A CD is a probability distribution that can quantify the uncertainty of an unknown parameter, where both
the bootstrap distribution in \citet{harris1989} and the fiducial distribution in \citet{wang2012} can be viewed as CDs;
see \citet{xie_singh_2013} for a review of these ideas.

To summarize, a predictive distribution can be constructed by using a data-based distribution on the parameter space to replace the posterior distribution in (\ref{bayespred}).
Following this idea, we aim to use draws from a joint probability distribution for the parameters such that the resulting predictive distribution can be used to construct asymptotically correct prediction bounds and intervals for within-sample prediction.
In particular, we propose two ways of constructing predictive distributions, extending the framework proposed by \citet{shen_liu_xie_2018} to the within-sample prediction case.
In Section~\ref{bootpred}, we describe a prediction method that is based on the bootstrap distribution
of the parameters and it is called the direct-bootstrap method in this paper. In Section~\ref{gpqpred},
we describe another method that works specifically with the (log)-location-scale family of distributions.
This method is inspired by generalized pivotal quantities (GPQ) and involves generating bootstrap samples
and it is called the GPQ-bootstrap method.

\subsection{The Direct-Bootstrap Method}
\label{bootpred}
For within-sample prediction, recall that number $Y_n$ of events between the censoring time $t_c$ and a future time $t_w>t_c$, given the Type~\Romannum{1} censored data $\boldsymbol{D}_n$, is $\mathrm{binomial}(n-r_n, p)$, where $r_n$ is the number of events
observed in $\boldsymbol{D}_n$ and $p$ is the conditional probability in (\ref{piequa}). 
The direct-bootstrap method uses the distribution of a bootstrap version $\widehat{p}_{n}^{*}=\pi(\bhtheta^\ast)$ of $\widehat{p}_{n}=\pi(\bhtheta)$, which is induced by the distribution of estimates $\bhtheta^\ast$ from a bootstrap sample $\boldsymbol{D}_n^\ast$,
to construct a predictive distribution.
Letting $\Pr_{\ast}$ denote bootstrap probability (probability induced by a bootstrap sample $\boldsymbol{D}^{*}_n$),
the predictive distribution constructed using direct-bootstrap method is
\begin{equation}
G_{Y_{n}}^{DB}(y|\boldsymbol{D}_n)=
\int\mathrm{pbinom}(y, n-r_n,\widehat{p}_n^\ast)\Pr{}_{\!\!*}\left(d \widehat{p}_{n}^{*}\right)
\approx\frac{1}{B} \sum_{b=1}^{B}\mathrm{pbinom}(y, n-r_n, \widehat{p}_b^\ast),
\label{bootpredformula}
\end{equation}
where $\widehat{p}_{1}^{*}, ...,\widehat{p}_{B}^{*}$ are realized bootstrap versions of $\widehat{p}_{n}$ from $B$ independently generated bootstrap samples $\boldsymbol{D}_n^{*(1)},\ldots,\boldsymbol{D}_n^{*(B)}$, and $B$ is the number of bootstrap samples.
The $100(1-\alpha)\%$ lower and upper prediction bounds using the direct-bootstrap method are then
\begin{equation}
\begin{split}
\utilde{Y}_{n, 1-\alpha}^{{DB}}&=\sup \left\{y \in \{0\}\cup\mathbb{Z}^{+}:G_{Y_{n}}^{DB}(y-1 | \boldsymbol{D}_n)\leq \alpha\right\},\\
\tilde{Y}_{n, 1-\alpha}^{{DB}}&=\inf \left\{y \in \{0\}\cup\mathbb{Z}^{+} :G_{Y_{n}}^{DB}(y | \boldsymbol{D}_n)\geq 1-\alpha\right\}.
\end{split}
\label{direct-bound}
\end{equation}

\subsection{The GPQ-Bootstrap Method}
\label{gpqpred}

This section focuses on the log-location-scale distribution family and develops another method to construct a predictive distribution through approximate GPQs.
Suppose $(T_1,..., T_n)$ is an i.i.d. random sample from a log-location-scale distribution
\begin{equation}
    F(t;\mu, \sigma)=\Phi\left[\frac{\log(t)-\mu}{\sigma}\right],
\label{log-location-scale}
\end{equation}
where $\Phi(\cdot)$ is a known cdf that is free of parameters.
For example, if $\Phi(\cdot)$ is the standard normal cdf $\Phi_{\mathrm{nor}(\cdot)}$, then $T_{1}$ has the log-normal distribution.

\citet{hannig2006} described methods for constructing GPQs and outlined the relationship between GPQs and fiducial inference. 
Applying these ideas, GPQs can be defined for the parameters $(\mu,\sigma)$ in the log-location-scale model as follows.
If $\mathbb{S}$ is a complete or Type~II censored independent sample from a log-location-scale distribution, a set of GPQs for $(\mu, \sigma)$ under $\mathbb{S}$ is given by
\begin{equation}
\begin{aligned}
\mu_n^{\ast\ast}=\widehat{\mu}_{n}+\left(\frac{\mu-\widehat{\mu}^{\mathbb{S}^{*}}_{n}}{\widehat{\sigma}^{\mathbb{S}^{*}}_{n}}\right) \widehat{\sigma}_{n} \quad\text{and}\quad \sigma^{\ast\ast}_n=\left(\frac{\sigma}{\widehat{\sigma}_{n}^{\mathbb{S}^{*}}}\right) \widehat{\sigma}_{n},\end{aligned}
\label{gpq}
\end{equation}
where $\mathbb{S}^{*}$ denotes an independent copy of the sample
$\mathbb{S}$, and $(\widehat{\mu}_{n}, \widehat{\sigma}_{n})$
and $(\widehat{\mu}^{\mathbb{S}^{*}}_{n}, \widehat{\sigma}^{\mathbb{S}^{*}}_{n})$
denote the ML estimators of $(\mu, \sigma)$
computed from $\mathbb{S}$ and $\mathbb{S}^{*}$, respectively.
These GPQs induce a distribution over the parameter space $(\mu,\sigma)$ based on data estimates $(\widehat{\mu}_n,\widehat{\sigma}_n)$ and, due to the fact that $[(\mu-\widehat{\mu}_n)/\sigma,\widehat{\sigma}_n/\sigma]$ are pivotal quantities based on a complete or Type~II censored sample $T_1,\dots,T_n$ from the log-location-family, the distribution of $[(\mu-\widehat{\mu}_n^{\mathbb{S}*})/\widehat{\sigma}_n^{\mathbb{S}*}, \sigma/\widehat{\sigma}_n^{\mathbb{S}*})]$ in (\ref{gpq}) can be directly approximated by simulation. 

GPQs can also, in some applications, be used to construct confidence
intervals when an exact pivot is unavailable.
Notice that, while the quantities in (\ref{gpq})
are GPQs for log-location-scale family
based on complete or Type~II censored data, these are no longer 
GPQs with Type~\Romannum{1} censored data,
where exact GPQs technically fail to exist.
This is because the distribution of
$\left[(\mu-\widehat{\mu}_{n})/\widehat{\sigma}_{n}, \sigma/\widehat{\sigma}_{n}\right]$
depends on the unknown event probability
$F(t_c;\mu,\sigma)$ before the censoring time $t_c$
under Type~\Romannum{1} censoring,
which applies also to
$\left[(\mu-\widehat{\mu}_{n}^{\mathbb{S}^{*}})/\widehat{\sigma}_{n}^{\mathbb{S}^{*}}, \sigma/\widehat{\sigma}_{n}^{\mathbb{S}^{*}}\right]$.

However, the formula in (\ref{gpq}) can be used to provide a joint approximate
GPQ distribution under Type~I censoring. Letting $\bhtheta^\ast = \left(\widehat{\mu}_{n}^{*}, \widehat{\sigma}_{n}^{*}\right)$ denote a bootstrap version of $\boldsymbol{\widehat{\theta}}_{n} = \left(\widehat{\mu}_{n}, \widehat{\sigma}_{n}\right)$,
(\ref{gpq}) is extended to define a joint approximate GPQ distribution as the bootstrap distribution of $\bhtheta^{\ast\ast} = \left(\widehat{\mu}_{n}^{**}, \widehat{\sigma}_{n}^{**}\right)$, where
\begin{equation}
\begin{aligned} \widehat{\mu}_{n}^{**} &=\widehat{\mu}_{n}+\left(\frac{\widehat{\mu}_{n}-\widehat{\mu}^{*}_{n}}{\widehat{\sigma}^{*}_{n}}\right) \widehat{\sigma}_{n}\quad\text{and}\quad\widehat{\sigma}^{**}_{n} &=\left(\frac{\widehat{\sigma}_{n}}{\widehat{\sigma}_{n}^{*}}\right) \widehat{\sigma}_{n}.\end{aligned}
\label{gpq2}
\end{equation}
The above definition of $\bhtheta^{**}$ also follows by using the bootstrap distribution of  $\left[(\widehat{\mu}_{n}-\widehat{\mu}_{n}^{*})/\widehat{\sigma}_{n}^{*}, \widehat{\sigma}_{n}/\widehat{\sigma}_{n}^{*}\right]$ to approximate the sampling distribution of $\left[(\mu-\widehat{\mu}_{n})/\widehat{\sigma}_{n}, \sigma/\widehat{\sigma}_{n}\right]$ and linearly solving for $(\mu,\sigma)$.
Then using $\bhtheta^{\ast\ast}=(\widehat{\mu}_n^{\ast\ast},\widehat{\sigma}_n^{\ast\ast})$ instead of $\bhtheta^\ast=(\widehat{\mu}_n^{*},\widehat{\sigma}_n^{*})$, a predictive distribution can be defined by
using the same procedure that defined the predictive distribution in (\ref{bootpredformula}).
Namely, by defining a random variable $\widehat{p}^{**}_{n}\equiv \pi(\bhtheta^{\ast\ast})$ from (\ref{piequa}) with a bootstrap distribution induced by $\bhtheta^{\ast\ast}=(\widehat{\mu}_{n}^{**}, \widehat{\sigma}_{n}^{**})$, the
predictive distribution for $Y_n$ using the GPQ-bootstrap method is given by
\begin{equation*}
    G_{Y_{n}}^{GPQ}(y | \boldsymbol{D}_n)=\int\mathrm{pbinom}(y, n-r_n, \widehat{p}_n^{**}) \Pr{}_{\!\!*}\left(d \widehat{p}_{n}^{**}\right)\approx\frac{1}{B}\sum_{b=1}^{B} \mathrm{pbinom}(y, n-r_n, \widehat p^{\ast\ast}_b),
\end{equation*}
where $\widehat p_1^{\ast\ast},\dots, \widehat p_B^{\ast\ast}$ are computed from realized bootstrap samples.
The $100(1-\alpha)\%$ lower and upper prediction bounds using GPQ-bootstrap method can be obtained
by replacing the predictive distribution $G_{Y_n}^{DB}(\cdot|\cdot)$ with $G_{Y_n}^{GPQ}(\cdot|\cdot)$
in (\ref{direct-bound}).

\subsection{Coverage Probability of the Proposed Methods}
This section shows that both the direct-bootstrap method (Section~\ref{bootpred})
and the GPQ-bootstrap method (Section~\ref{gpqpred})
produce asymptotically correct prediction bounds/intervals
for the future count $Y_n$.
Hence, these two methods yield asymptotically valid inference
for within-sample prediction of $Y_n$,
as does the calibration-bootstrap method (Theorem~\ref{theocali}, Section~\ref{calibration}),
but not by the standard plug-in method (Theorem~\ref{first_theorem}, Section~\ref{plugin_not_regular}).

\begin{theorem}
Under the same conditions as Theorem~\ref{theocali},
\begin{enumerate}
\item The $100(1-\alpha)\%$ upper and lower prediction bounds using the direct-bootstrap method,
respectively $\tilde{Y}^{DB}_{n, 1-\alpha}$ and $\utilde{Y}^{DB}_{n, 1-\alpha}$, have asymptotically correct coverage. That is,$$\lim_{n\rightarrow\infty}\Pr(Y_{n}\leq\tilde{Y}^{DB}_{n, 1-\alpha}) = 1-\alpha=\lim_{n\rightarrow\infty}\Pr(Y_n\geq\utilde{Y}_{n, 1-\alpha}^{DB}).$$
\item If the parametric distribution $F(\cdot; \mu, \sigma)$ belongs to the log-location-scale distribution family (\ref{log-location-scale}), with standard cdf $\Phi(\cdot)$ differentiable on $\mathbb{R}$, the $100(1-\alpha)\%$ upper and lower prediction bounds using the GPQ-bootstrap method, respectively $\tilde{Y}^{GPQ}_{n, 1-\alpha}$ and $\utilde{Y}^{GPQ}_{n,1-\alpha}$, have asymptotically correct coverage. That is, $$\lim_{n\rightarrow\infty}\Pr(Y_{n}\leq\tilde{Y}^{GPQ}_{n, 1-\alpha}) = 1-\alpha=\lim_{n\rightarrow\infty}\Pr(Y_n\geq\utilde{Y}_{n,1-\alpha}^{GPQ}).$$
\end{enumerate}
\label{predbound}
\end{theorem}
The proof of Theorem~3 is in the online supplementary material.

\section{Multiple Cohort Within-Sample Prediction}
\label{sec:multiple-cohort}
\subsection{Multiple Cohort Data}
So far, the focus has been on the within-sample prediction for single-cohort data.
Multiple cohort data, however, are more common in applications.
In this section, the results from single-cohort data are extended to multiple-cohort data.

In multiple-cohort data (e.g. the bearing cage data of Section~\ref{m_exampls}),
units from different cohorts are placed into service at different times.
The multiple-cohort data $\mathbb{D}$ can be seen as a collection of several
single-cohort datasets as $\mathbb{D}=\{\boldsymbol{D}_{n_{s}}, s=1,...,S\}$,
where $S$ is the number of cohorts and $n_s$ is the number of units in the cohort $s$
(sometimes, with no grouping, many cohorts have size 1).
Within each cohort $\boldsymbol{D}_{n_{s}}=(D_{s,1},...,D_{s,n_s})$,
we may express an observation involved as $D_{s,i}=(\delta_i^s, T^{obs,s}_{i})$,
for $T^{obs,s}_{i}=T_i^s\delta_i^s+(1-\delta_i^s)t_c^s$,
where $T_i^s$ is a random variable from a parametric distribution $F(\cdot;\btheta)$,
$t_c^s$ is the censoring time for cohort $s$,
and $\delta_i^s=\mathrm{I}(T_i^s\leq t_c^s)$ is a random variable
indicating whether a unit's value (e.g., failure time)
is less than the censoring time $t_c^s$.
Given the multiple-cohort data $\mathbb{D}$,
the number of observed events (e.g., failures)
within cohort $s$ is defined as
$r_{n_s}=\sum_{i=1}^{n_s}\mathrm{I}(T_i^s\leq t_c^s),s=1,...,S$,
where the total number of units is $n=\sum_{s=1}^{S}n_s$.
The predictand in the multiple-cohort data is the total number
of events that will occur in a future time window of length $\Delta$
and it is denoted by
$Y_n=\sum_{s=1}^{S}\sum_{i=1}^{n_s}\mathrm{I}(t_c^s<T^s_i\leq t_w^s)$,
where $t_w^s=t_c^s+\Delta$ for $s=1,\dots, S$.

Within each cohort $s=1,...,S$,
the number $Y_s =\sum_{i=1}^{n_s}\mathrm{I}(t_c^s < T_i^s \leq t_w^s)$
of future events has a binomial distribution.
As in Section 3, the conditional distribution of $Y_s$
is $\mathrm{binomial}(n-r_{n_s}, p_s)$, where $p_s$ is defined as
\begin{align*}
    p_s\equiv\pi_{s}(\btheta)=\frac{F(t_w^s;\btheta)-F(t_c^s;\btheta)}{1-F(t_c^s;\btheta)},\quad s=1,\dots,S.
\end{align*}
Consequently, the predictand $Y_n=\sum_{s=1}^{S}Y_s$ has
a Poisson-binomial distribution with probability vector
$\boldsymbol{p}=(p_1,...,p_S)$ and weight vector
$\boldsymbol{w}=(n_1-r_{n_1},...,n_S-r_{n_S})$.
We denote this Poisson-binomial distribution by
$\mathrm{Poibin(\boldsymbol{p}, \boldsymbol{w})}$,
where the cdf of the Poisson-binomial distribution is denoted by
$\mathrm{ppoibin}(\cdot, \boldsymbol{p}, \boldsymbol{w})$
and the quantile function is denoted by $\mathrm{qpoibin(\cdot,
\boldsymbol{p}, \boldsymbol{w})}$; these functions are
available in the \textbf{poibin} R package (described in \citet{hongpoisson2013}).

If $\bhtheta$ is a consistent estimator of $\btheta$ based on the multiple-cohort data $\mathbb{D}$, an estimator $\widehat{\boldsymbol{p}}=(\widehat p^1_{n},... ,\widehat p^S_{n})$ of conditional probabilities $\boldsymbol{p}$ follows by substitution $\widehat{p}_n^s = \pi_s(\bhtheta)$, $s=1,\ldots,S$, similar to the single-cohort case. Then, the $100(1-\alpha)\%$ plug-in lower and upper prediction bounds for $Y_n$ are
\begin{align*}
\utilde{Y}^{PL}_{n, 1-\alpha}&=
\sup \{ y\in \{0\}\cup\mathbb{Z}^{+}: \mathrm{ppoibin}\left(y-1, \widehat{\boldsymbol{p}}, \boldsymbol{w}\right) \leq\alpha\}\\
&=\begin{cases}
\mathrm{qpoibin}(\alpha, \widehat{\boldsymbol{p}}, \boldsymbol{w}), &\text{if } \mathrm{pbinom}(\mathrm{qpoibin}(\alpha, \widehat{\boldsymbol{p}}, \boldsymbol{w}), \widehat{\boldsymbol{p}}, \boldsymbol{w})>\alpha,\\
\mathrm{qpoibin}(\alpha, \widehat{\boldsymbol{p}}, \boldsymbol{w})+1, &\text{if } \mathrm{pbinom}(\mathrm{qpoibin}(\alpha, \widehat{\boldsymbol{p}}, \boldsymbol{w}), \widehat{\boldsymbol{p}}, \boldsymbol{w})=\alpha,
\end{cases}\\
\tilde{Y}_{n,1-\alpha}^{PL}&=\inf\{y\in\{0\}\cup\mathbb{Z}^{+}:\mathrm{ppoibin}(y, \widehat{\boldsymbol{p}}, \boldsymbol{w})\geq1-\alpha\}=\mathrm{qpoibin}(1-\alpha, \widehat{\boldsymbol{p}}, \boldsymbol{w}).
\end{align*}
Similar to the single-cohort case (Theorem~\ref{first_theorem}), the plug-in method also fails to provide an asymptotically correct coverage probability under multiple-cohort data;
see the online supplementary material.

\subsection{The Calibration-Bootstrap Method for Multiple Cohort Data}
\label{calibration_multiple_cohort_data}
Formulating prediction bounds using the calibration-bootstrap method
first requires simulation of bootstrap samples,
where each bootstrap sample $\mathbb{D}^\ast$ matches the original data in terms of the number
$S$ of cohorts as well as their respective sizes $n_s$ and censoring times $t_c^s$, $s=1,\ldots,S$. 
The bootstrap version of the estimator $\widehat{\boldsymbol{p}}=(\widehat p^1_{n},... ,\widehat p^S_{n})$ is $\widehat{\boldsymbol{p}}^{\ast}=(\widehat p^{1,\ast}_{n},... ,\widehat p^{S,\ast}_{n})$ from each bootstrap sample $\mathbb{D}^*$.
Additionally, the number of events (e.g., failures) in the bootstrap sample, grouped by cohort,
is $(r_{n_1}^{\ast},...,r_{n_S}^{\ast})$,
from which we denote a bootstrap future count by
$Y_n^{\dagger}\sim\text{Poibin}(\widehat{\boldsymbol{p}}; \boldsymbol{w}^\ast)$
based on a weight vector from the bootstrap sample as
$\boldsymbol{w}^{\ast}=(n_1-r_{n_1}^{\ast},...,n_S-r_{n_S}^{\ast})$.
The bootstrap variable set
$(Y_n^\dagger, \widehat{\boldsymbol{p}}^{\ast}, \boldsymbol{w}^{\ast})$
is then applied into a Poisson-binomial cdf and then leads to a transformed random variable
$U^\ast=\mathrm{ppoibin}(Y_{n}^\dagger, \widehat{\boldsymbol{p}}^{\ast}, \boldsymbol{w}^\ast)\in[0,1]$
for deriving calibrated confidence levels
$\alpha^\dagger_L$ and $\alpha^\dagger_U$
in the same way as in the single-cohort situation.
Then, the $100(1-\alpha)\%$ calibrated lower prediction bound is
$\utilde{Y}^{C}_{n,1-\alpha}=\utilde{Y}^{PL}_{n,1-\alpha^\dagger_L}$
and the similar upper prediction bound version is
$\tilde{Y}^{C}_{n,1-\alpha}=\tilde{Y}^{PL}_{n,1-\alpha^\dagger_U}$.

The calibration-bootstrap method remains asymptotically correct
for multiple-cohort within-sample prediction.
The multiple-cohort extensions of Theorem~\ref{theocali}
and the algorithm are in the online supplementary material.

\subsection{The Direct- and GPQ-Bootstrap Methods for Multiple Cohort Data}
For multiple-cohort data, constructing prediction bounds
for $Y_n$ based on the predictive-distribution-based methods
also requires bootstrap data and, in particular,
the distribution of a bootstrap version
$\widehat{\boldsymbol{p}}^\ast$ of $\widehat{\boldsymbol{p}}$
as in Section~\ref{calibration_multiple_cohort_data}.
The predictive distribution from the direct-bootstrap method is
\begin{equation}\label{bootppoi}
G^{DB}_{Y_n}(y|\mathbb{D})=\int \mathrm{ppoibin}(y, \widehat{\boldsymbol{p}}^{\ast}, \boldsymbol{w})\Pr{}_{\!\!\ast}(d \widehat{\boldsymbol{p}}^{\ast})\approx\frac{1}{B}\sum_{b=1}^{B}\mathrm{ppoibin}(y, \widehat{\boldsymbol{p}}^\ast_b, \boldsymbol{w}).
\end{equation}
where $\widehat{\boldsymbol{p}}^\ast_1,\dots, \widehat{\boldsymbol{p}}^\ast_B$ are realized bootstrap versions of $\widehat{\boldsymbol{p}}$ across independently generated bootstrap versions of multiple-cohort data (e.g., $\mathbb{D}^*$).
The $100(1-\alpha)\%$ direct-bootstrap lower and upper prediction bounds for $Y_n$ are defined as the modified $\alpha$ quantile and $1-\alpha$ quantile of this predictive distribution, respectively, and given by
\begin{align*}
\utilde{Y}_{n, 1-\alpha}^{DB}&=\sup \left\{y \in\{0\} \cup \mathbb{Z}^{+}: G_{Y_{n}}^{DB}\left(y-1 | \mathbb{D}\right) \leq\alpha\right\},\\
\tilde{Y}_{n, 1-\alpha}^{DB}&=\inf \left\{y \in\{0\} \cup \mathbb{Z}^{+}: G_{Y_{n}}^{DB}\left(y | \mathbb{D}\right) \geq 1-\alpha\right\}.
\end{align*}

If $F(\cdot;\btheta)=F(\cdot;\mu,\sigma)$ belongs to the log-location-scale family as in (\ref{log-location-scale}),
we use $\widehat{\boldsymbol{\theta}}_n^{\ast}=(\hat\mu_n^\ast,\hat\sigma_n^\ast)$ to compute approximate GPQs
$\widehat{\boldsymbol{\theta}}_n^{\ast\ast}=(\hat\mu_n^{\ast\ast},\hat\sigma_n^{\ast\ast})$ using (\ref{gpq2}),
and compute $\widehat{\boldsymbol{p}}^{\ast\ast}=(\widehat p_{n}^{1,\ast\ast},\dots,\widehat p_{n}^{S,\ast\ast})$ where $\widehat p_{n}^{s, \ast\ast}=\pi_s(\widehat{\boldsymbol{\theta}}^{\ast\ast}_n)$.
Then the GPQ-bootstrap method can be implemented to obtain prediction bounds for $Y_n$ by replacing $\widehat{\boldsymbol{p}}^\ast$ with $\widehat{\boldsymbol{p}}^{\ast\ast}$ in the definition of the direct-bootstrap predictive distribution (\ref{bootppoi}) and analogously determining prediction bounds from the quantiles of this predictive distribution.
The direct- and GPQ-bootstrap methods produce asymptotically correct prediction bounds from multiple-cohort data, and the extension of Theorem~\ref{predbound} is provided in the online supplementary material.


\section{A Simulation Study}
\label{simu:study}
The purpose of this simulation study is to illustrate agreement
for finite sample sizes with the theorems established in the
previous sections and to provide insights into the performance
of different methods in the case of finite samples.
The details and results in this section are for Type~\Romannum{1} censored single-cohort data.
Let the event of interest be the failure of a unit.
We simulated Type~\Romannum{1} censored data using the two-parameter Weibull distribution and compared the coverage probabilities of the prediction bounds based on the plug-in, calibration-bootstrap, direct-bootstrap, and GPQ-bootstrap methods. The Weibull cdf is
$$
F(t;\eta, \beta) = 1-\exp\left[-\left(\frac{t}{\eta}\right)^{\beta}\right],\quad t>0,
$$
with positive scale $\eta$ and shape $\beta$ parameters, and can also be parameterized as
$$
F(t;\mu, \sigma) = \Phi_{\textrm{sev}}\left[\frac{\log (t)-\mu}{\sigma}\right],\quad t>0,
$$
where $\Phi_{\textrm{sev}}(x)=1-\exp\left[-\exp(x)\right]$ is the cdf of the standard smallest extreme value distribution with $\mu = \log (\eta)$ and $\sigma = 1/\beta$.
The conditions in Theorems~1-3 can be verified for Type~I censored Weibull data, so that the Weibull distribution can be used to illustrate all of the aforementioned methods for within-sample prediction (e.g., the ML estimators of the Weibull parameters $\bhtheta = (\widehat{\mu}_n,\widehat{\sigma}_n)$ have sampling distributions with normal limits and can be validly approximated by parametric bootstrap as described
in \citet{scholz1996maximum}).


\subsection{Simulation Setup}
The factors for the simulation experiment are
(i) $p_{f1} = F(t_c;\beta,\eta)$, the probability that a unit fails before the censoring time $t_{c}$;
(ii) $\mathrm{E}(r)= np_{f1}$, the expected number of failures at the censoring time $t_c$, where $n$ is the total sample size (i.e., including both the censored and the uncensored observations);
(iii) $d \equiv p_{f2}-p_{f1}$, the probability that a unit fails in a future time interval $(t_c,t_w]$ where $p_{f2} = F(t_w;\beta, \eta)$;
(iv) $\beta = 1/\sigma$, the Weibull shape parameter.
Because $\eta=\exp(\mu)$ is a scale parameter, without loss of generality, $\eta=1$ was used in the simulation.
A simulation with all combinations of the following factors levels was conducted:
(i) $p_{f1} = 0.05, 0.1, 0.2$;
(ii) $\mathrm{E}(r) = 5, 15, 25, 35, 45$;
(iii) $d = 0.1, 0.2$;
(iv) $\beta = 0.5, 0.8, 2, 4$.

For each combination of the these four factors, 90\% and 95\% upper prediction bounds and 90\% and 95\% lower prediction bounds were constructed.

The procedure for the simulation is as follows:
\begin{enumerate}
	\itemsep-0.5em
	\item Simulate $N$=5000 Type~\Romannum{1} censored samples for each of the factors-level combinations of the four factors.
	\item Use ML to estimate parameters $\beta,\eta$ in each censored sample.
	\item  Compute prediction bounds using the different methods for each sample.
	\item Compute the conditional (i.e., binomial) coverage probability for each of the prediction bounds.
	\item Determine the unconditional coverage probability for each method by averaging the $N=5000$ conditional coverage probabilities.
\end{enumerate}

Within each of the $N$=5000 simulated Type~I censored samples, $B$=5000  bootstrap samples were generated by parametric bootstrap (i.e., as a random sample from the fitted Weibull distribution with Type~I censoring at $t_c$) and these
samples were used for the calibration-bootstrap method and the two predictive-distribution-based methods.
In the simulation, we excluded those samples having fewer than 2 failures to avoid estimability problems, so that all $N$=5000 original samples and all the $N\times B$=25{,}000{,}000 bootstrap samples in the simulation have at least 2 failures.
The probability of a data sample with fewer than 2 failures for each factor-level combination is given in Table~\ref{table:droprate}.

\begin{table}[ht!]
\centering
\begin{tabular}{cccccc}
\hline
     & {E}($r$)=5 & {E}($r$)=15 & {E}($r$)=25 & {E}($r$)=35 & {E}($r$)=45\\\hline
    $p_{f1}=0.05$ & 0.037 &0.000&0.000&0.000&0.000\\
    $p_{f1}=0.1$ & 0.034 &0.000&0.000&0.000&0.000\\
    $p_{f1}=0.2$& 0.027 &0.000&0.000&0.000&0.000\\\hline
\end{tabular}
\caption{Probability of an excluded sample (i.e., $r=0$ or 1 failures) for different factor-level combinations.}
\label{table:droprate}
\end{table}

\subsection{Simulation Results}

\setcounter{figure}{0}
\makeatletter 
\renewcommand{\thefigure}{\arabic{figure}}
A small subset of the plots displaying the complete simulation
results are given here, as the results are generally consistent
across the different factor-level combinations.
Figure~\ref{threeMethods} shows
the coverage probabilities from plug-in, calibration-bootstrap, direct-bootstrap,
and GPQ-bootstrap methods when $\beta = 2$ and $d = 0.2$.
The horizontal dashed line in each subplot represents
the nominal confidence level.
Plots for the other factor-level combinations are given in the online supplementary material.
\begin{figure}[t]
\centering
\includegraphics[width=\textwidth]{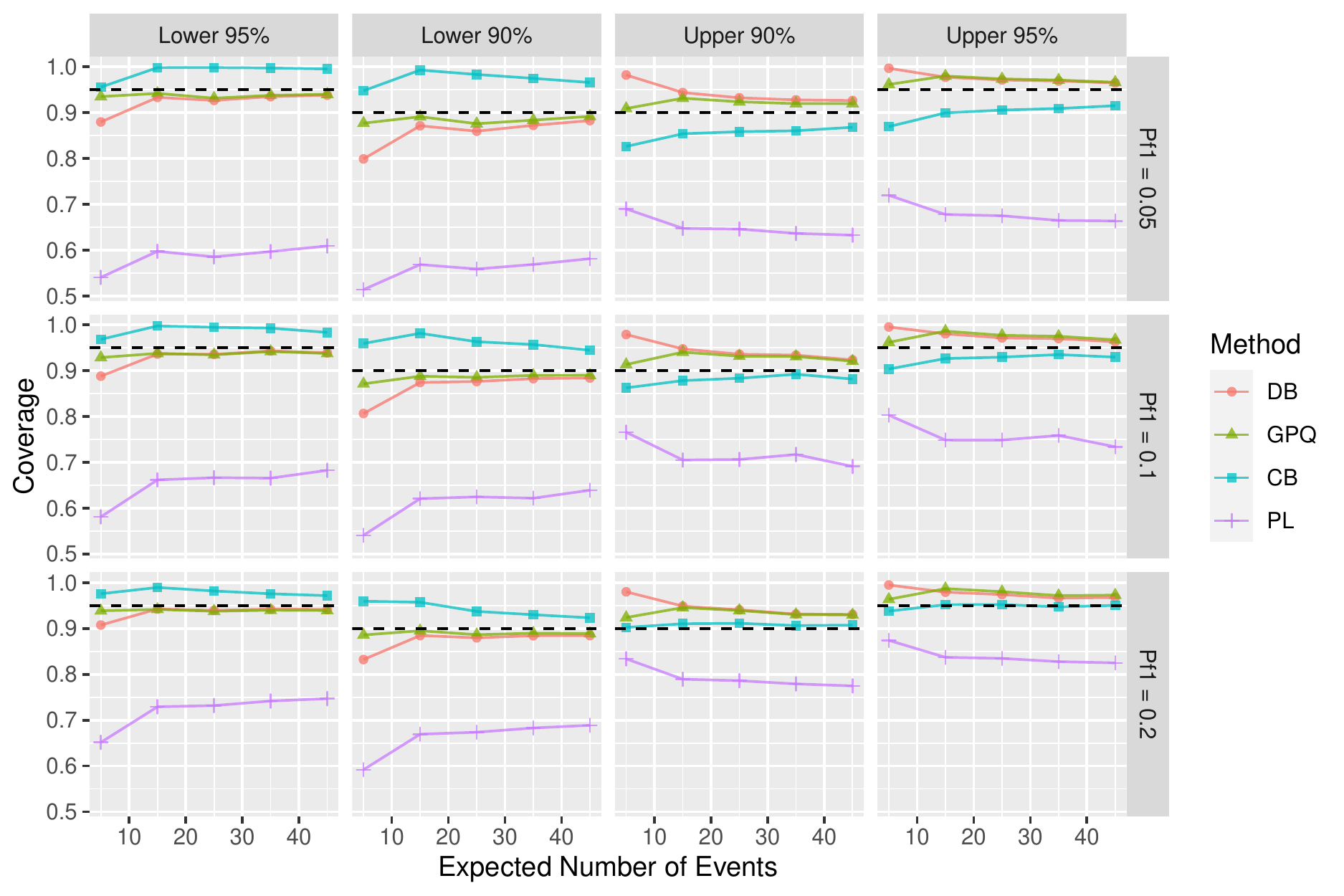}
\caption{Coverage probabilities versus expected number of events for the direct-bootstrap (DB),
	GPQ-bootstrap (GPQ), calibration-bootstrap (CB), and plug-in (PL) methods when $d=p_{f2}-p_{f1}=0.2$ and $\beta = 2$.}
\label{threeMethods}
\end{figure}

Some observations from the simulation results are:
\begin{enumerate}
	\itemsep-0.5em
\item The plug-in method fails to have asymptotically correct coverage probability.
As $p_{f1}$ decreases, which entails less information or fewer events observed before the censoring time $t_c$, the coverage probability deviates more from the nominal level.
\item The direct- and GPQ-bootstrap methods are close to each other in terms of coverage probabilities except when $\mathrm{E}(r)=5$. The calibration-bootstrap method differs considerably from the direct- and GPQ-bootstrap methods.
The calibration-bootstrap method tends to be more conservative than the other bootstrap-based methods for constructing lower prediction bounds,
and also is less conservative for constructing upper prediction bounds.
\item For the lower bounds, the direct- and GPQ-bootstrap methods dominate the calibration-bootstrap method.
For the upper bounds, the coverage probabilities of
the former two bootstrap-based methods
are slightly conservative but still close to the nominal level.
The calibration-bootstrap method is better than the direct- and GPQ-bootstrap methods in just a few of these upper bounds.
\item Compared with the calibration-bootstrap method,
whose performance is highly related to the level of $p_{f1}$, the coverage probabilities of the direct- and GPQ-bootstrap methods are insensitive to the level of $p_{f1}$.
As $p_{f1}$ decreases, the lower prediction bound using the calibration-bootstrap method has over-coverage while the upper prediction bound has under-coverage.
This implies that under heavy censoring (small $p_{f1}$), extremely large
sample sizes $n$ (or correspondingly large expected number of failing $\text{E}(r)=n p_{f1}$)
are required to attain coverage probabilities close to the nominal confidence level.
\end{enumerate}

From these observations, we can see that the direct- and GPQ-bootstrap methods (i.e., predictive-distribution-based methods) tend to dominate the calibration-bootstrap method in terms of the performance of the prediction bounds, even though all three methods are asymptotically valid.
This is because the predictive-distribution-based methods target the one source $p$ of parameter uncertainty in the conditional $\text{binomial}(n-r_n,p)$ distribution of the predictand $Y_n$ (i.e., as addressed by applying bootstrap versions $\widehat{p}^\ast$ or $\widehat{p}^{\ast\ast}$ to ``smooth'' estimation uncertainty for $p$), while the number $n-r_n$ of Bernoulli trials used in these predictive distributions matches that of the predictand.
Due to its definition, however, the calibration-bootstrap method involves bootstrap approximation steps (i.e., $r^*_n, \widehat{p}^*$) for both the number $r_n$ of failures as well as the binomial probability $p$.
The calibration-bootstrap method essentially imposes an approximation $n-r^*_n$ for the known number
$n-r_n$ of trials prescribing the predictand $Y_n$.
As a consequence, coverages from the calibration-bootstrap method are generally
less accurate than those from the predictive-distribution-based methods for within-sample prediction.

\section{Application of the Methods}
\label{sec:applications}
\subsection{Examples}
\noindent \textbf{Product-A Data}:
The ML estimates of the Weibull shape and scale parameters are $\widehat\beta=1.518$ and $\widehat\eta=1152$, respectively, based on 80 failure times among 10,000 units before 48 months.
Then, for the 9920 surviving units, the ML estimate of the probability that a unit will fail between 48 and 60 months of age is
$
\widehat p_n = [F(60;\widehat\beta, \widehat\eta)-F(48;\widehat\beta, \widehat\eta)]/[1-F(48;\widehat\beta, \widehat\eta)]= 0.00323.
$
Using the ML estimates of the Weibull parameters $(\widehat\beta, \widehat\eta)$, we simulate 10,000 bootstrap samples that are censored at 48 months and obtain ML estimates of $(\beta, \eta)$ from each bootstrap sample.
Based on applying these with each interval method, Table~\ref{productAData} gives prediction bounds for the number of failures in the next 12 months.
As indicated by our results, even with a large number of failures, the plug-in method intervals can be expected to be off and are too narrow compared to the other bounds.

\begin{table}[!ht]
\centering
\begin{tabular}{c c c c c c} 
\hline
Confidence Level &Bound Type& Plug-in & Direct & GPQ & Calibration \\ [0.5ex] 
 \hline
 95\% & Lower &\multicolumn{1}{r}{23} & \multicolumn{1}{r}{20} & \multicolumn{1}{r}{20} & \multicolumn{1}{r}{20} \\ 
 90\% & Lower &\multicolumn{1}{r}{25} & \multicolumn{1}{r}{23} & \multicolumn{1}{r}{23} & \multicolumn{1}{r}{23} \\
 90\% & Upper &\multicolumn{1}{r}{39} & \multicolumn{1}{r}{43} & \multicolumn{1}{r}{43} & \multicolumn{1}{r}{43} \\
 95\% & Upper &\multicolumn{1}{r}{42} & \multicolumn{1}{r}{47} & \multicolumn{1}{r}{47} & \multicolumn{1}{r}{46} \\
 \hline
\end{tabular}
\caption{Product A Data: Prediction Bounds for the number of failures in the next 12 months using different methods.}
\label{productAData}
\end{table}

\noindent \textbf{Heat Exchanger Data}: In this example, there are no exact failure times in the data.
That is, the data here contain limited information as there were only 8 failures among 20,000 exchanger tubes that were inspected (in censored data analysis, the informational content of data is closely related to the number of failures) and these failure times are interval-censored (not exact).
The likelihood function under a Weibull model for the heat exchanger data is
$$
L(\beta, \eta)=F(1; \beta, \eta)[F(2; \beta, \eta)-F(1; \beta, \eta)][F(3; \beta, \eta)-F(2; \beta, \eta)]^{6}[1-F(3; \beta, \eta)]^{19992},
$$
resulting in ML estimates $\widehat\beta=2.531$ and $\widehat\eta=66.058$.
The conditional probability of a tube failing between the third and tenth year, given that tube has not failed at the end of the third year, is then estimated as $\widehat p_n = [F(10;\widehat\beta, \widehat\eta)-F(3;\widehat\beta, \widehat\eta)]/[1-F(3;\widehat\beta, \widehat\eta)]= 0.00797$.

\begin{figure}[ht!]
\centering
\includegraphics[width=0.9\textwidth]{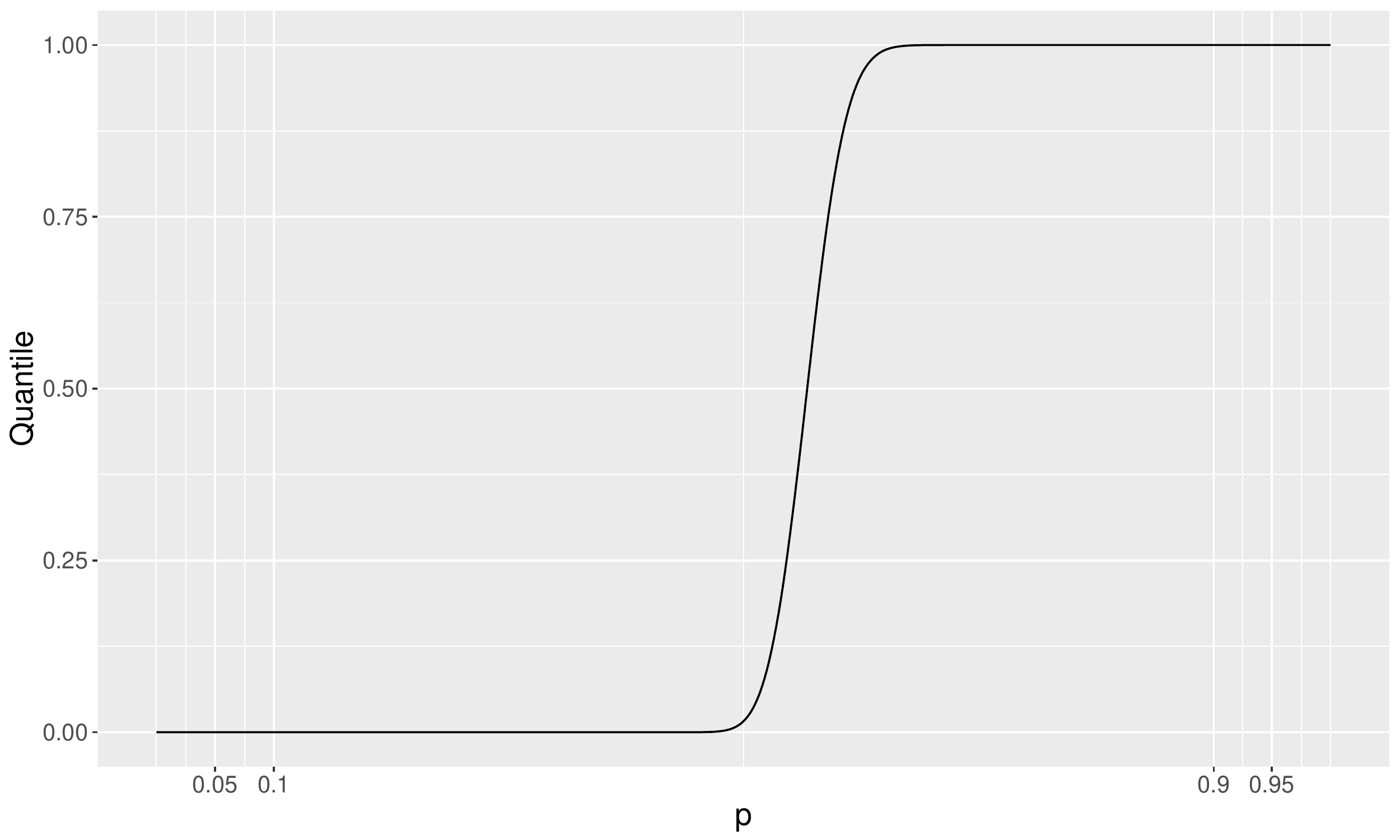}
\caption{The quantile function of $\mathrm{pbinom}(Y^\dagger_n, n-r^\ast_n, \widehat p^\ast_n)$ used for the calibration-bootstrap method with heat exchanger data.}
\label{calibrationquantile}
\end{figure}

The ML estimates from 10,000 bootstrap samples
(parametric bootstrap with censoring at 3 years)
are used in the calibration-bootstrap and two predictive-distribution-based methods.
However, the calibration-bootstrap method exhibits
numerical instabilities with these data due to the small number of failures.
To illustrate, Figure~\ref{calibrationquantile} shows the approximate quantile
function of $U^\ast=\mathrm{pbinom}(Y_n^\dagger, n-r_n^\ast, \widehat p_n^\ast)$
used in the calibration-bootstrap method, involving the evaluation of a $\text{binomial}(n-r^\ast_n,\widehat{p}_n^*)$ random variable $Y_n^\dagger$ in its cdf $\mathrm{pbinom}$, given the number $r^*_n$ of failures
and the estimate $\widehat{p}_n^*$ from a bootstrap sample.
This quantile function is also the calibration curve,
where the x-axis gives the desired confidence level $1-\alpha$,
while the y-axis gives the corresponding calibrated confidence
level ($\alpha^\dagger_L$ or $1-\alpha^\dagger_U$) to be used
for determining plug-in prediction bounds (or quantiles from a
$\text{binomial}(n-r_n=19992,\widehat{p}=0.00797)$ distribution).
From Figure~\ref{calibrationquantile}, we can see that the $0.05$ and $0.1$ quantiles nearly equal $0$
while the $0.9$ and $0.95$ quantiles nearly equal 1.
This creates complications in computing the prediction bounds,
for example, as there is numerical instability near the 100\% quantile of the $\text{binomial}(n-r_n=19992,\,\widehat{p}=0.00797)$ distribution.
Consequently, 90\% and 95\% bounds from the calibration-bootstrap
method are computationally not available (NA).
\begin{table}[ht]
\centering
\begin{tabular}{c c c c c c} 
\hline
 Confidence Level& Bound Type & Plug-in & Direct & GPQ & Calibration \\ [0.5ex] 
 \hline
 95\% & Lower &\multicolumn{1}{r}{138} & \multicolumn{1}{r}{28} & \multicolumn{1}{r}{23} & \multicolumn{1}{r}{NA}\\ 
 90\% & Lower &\multicolumn{1}{r}{142} & \multicolumn{1}{r}{43} & \multicolumn{1}{r}{34} & \multicolumn{1}{r}{NA}\\
 90\% & Upper &\multicolumn{1}{r}{176} & \multicolumn{1}{r}{1627} & \multicolumn{1}{r}{888} & \multicolumn{1}{r}{NA}\\
 95\% & Upper &\multicolumn{1}{r}{180} & \multicolumn{1}{r}{4343} & \multicolumn{1}{r}{1890} & \multicolumn{1}{r}{NA}\\
 \hline
\end{tabular}
\caption{Heat Exchanger Data: Prediction Bounds for the number of failures in the next 7 years using different methods.}
\label{heatExchangerData}
\end{table}
Table~\ref{heatExchangerData} instead provides prediction
bounds from the plug-in and direct- and GPQ-bootstrap methods. The plug-in prediction bounds
differ substantially from the two bootstrap-based methods.
Unlike the previous example (Product A data),
the direct- and GPQ-bootstrap methods also differ appreciably
based on the limited failure information with the heat exchanger data;
we return to explore such differences in Section~\ref{compare:gpq:boot}.
The upper bounds involve a large amount of extrapolation
and may not be practically meaningful other than to warn
that there is a huge amount of uncertainty in the
10-year predictions.

\noindent \textbf{Bearing Cage Data}: In this example,
staggered entry data containing multiple cohorts are considered.
Table~{\ref{bearingcage}} gives the prediction bounds for the bearing cage dataset using 10,000 bootstrap samples.
While similar in spirit to the Product-A
example, the predictand here differs by having a
Poisson-binomial distribution.
The latter can be computed with the R package \textbf{poibin},
which is applied to construct prediction bounds using methods
described in Section~\ref{calibration_multiple_cohort_data}.
Table~\ref{bearingcage} gives the resulting prediction bounds for the bearing
cage dataset.
\begin{table}[ht]
\centering
\begin{tabular}{c c c c c c} 
\hline
 Confidence Level & Bound Type & Plug-in & Direct & GPQ & Calibration \\ [0.5ex] 
\hline
 95\% & Lower &\multicolumn{1}{r}{2} & \multicolumn{1}{r}{1} & \multicolumn{1}{r}{1} & \multicolumn{1}{r}{1} \\ 
 90\% & Lower &\multicolumn{1}{r}{2} & \multicolumn{1}{r}{2} & \multicolumn{1}{r}{2} & \multicolumn{1}{r}{2} \\
 90\% & Upper &\multicolumn{1}{r}{8} & \multicolumn{1}{r}{10} & \multicolumn{1}{r}{13} & \multicolumn{1}{r}{10} \\
 95\% & Upper &\multicolumn{1}{r}{9} & \multicolumn{1}{r}{12} & \multicolumn{1}{r}{20} & \multicolumn{1}{r}{12} \\
 \hline
\end{tabular}
\caption{Bearing Cage Data: Prediction Bounds for the number of failures in the next 300 service hours using different methods.}
\label{bearingcage}
\end{table}

\subsection{Comparing the Direct- and GPQ-Bootstrap Methods}
\label{compare:gpq:boot}
In the heat exchanger example, the prediction bounds obtained from the
direct- and GPQ-bootstrap methods appear very different.
This motivates us to investigate the cause of such differences in similar prediction applications involving limited information.

A general simulation setting is first described for
mimicking the heat exchanger data.
The heat exchanger data has two important features in that
the number of events is small (i.e., 8) and so is the
proportion of observed events (i.e., 0.004).
Hence, in the simulation, the expected number of events $\text{E}(r)$
is set to 5 while the proportion failing $p_{f1}$
is 0.001, with a Weibull shape parameter $\beta=2$ and scale parameter $\eta = 1$.
Different levels of $d = p_{f2}-p_{f1}$ are used for the probability of
events in the forecast window.
The simulation results (available in the online supplementary material)
reveal that, overall, the GPQ-bootstrap method has better coverage probability
than the direct-bootstrap method in this simulation setting.
For the upper prediction bound, the direct-bootstrap method is
generally more conservative than the GPQ-bootstrap method in terms
of coverage probability, indicating that upper prediction bounds
from the direct-bootstrap method are larger than the GPQ counterparts.
On the other hand, the lower bound based on the direct-bootstrap method
generally tends to have under-coverage compared to the GPQ-bootstrap method,
suggesting also larger lower bounds from the direct-bootstrap method
relative to the GPQ-bootstrap method.
These patterns in the prediction bounds (i.e., with larger direct-bootstrap
bounds compared to those from the GPQ-bootstrap in a setting of a limited number of events) are consistent with the prediction bounds found
from the heat exchanger example.
\begin{figure}[ht!]
\centering
\includegraphics[width=0.85\textwidth]{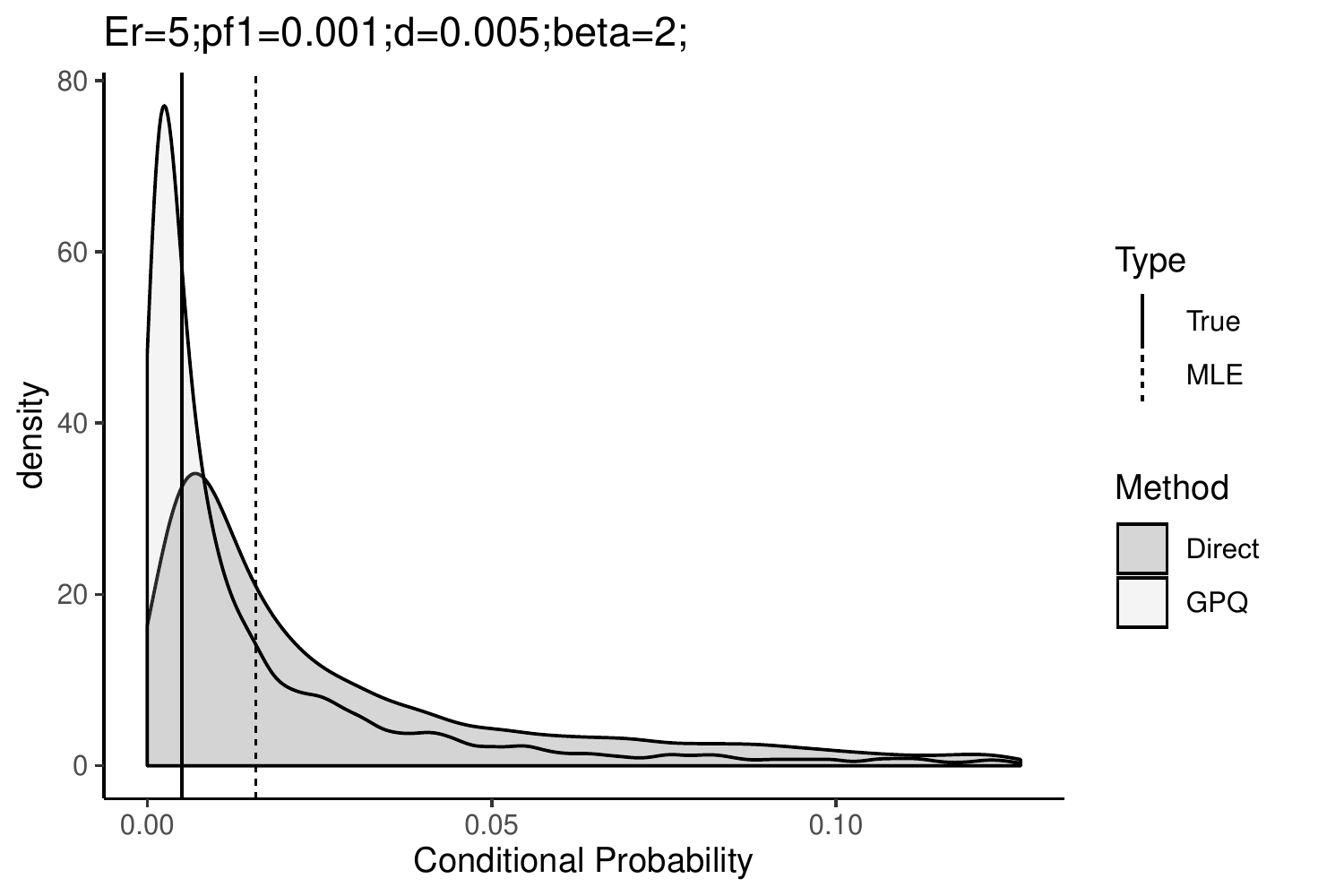}
\caption{A Representative Distribution of $\widehat p^{\ast}$ and $\widehat p^{\ast\ast}$.}
\label{fig:2}
\end{figure}
To further illustrate, Figure~\ref{fig:2} shows the
bootstrap distributions of $\widehat{p}^*$ and $\widehat{p}^{**}$ from
a single Monte Carlo sample that represents the typical behavior
found in this simulation setting:
values of $\widehat{p}^{**}$ used in the predictive distribution of GPQ-bootstrap method
tend to be smaller and more concentrated than the $\widehat{p}^*$ values
used in the direct-bootstrap predictive distribution.
Note that direct- and GPQ-bootstrap predictive distributions are approximated by
$G^{DB}_{Y_n}(y|\boldsymbol{D}_n)\approx1/B\sum_{b=1}^{B}\text{pbinom}(y, n-r_n, \widehat p^\ast_b)$
and $G^{GPQ}_{Y_n}(y|\boldsymbol{D}_n)\approx1/B\sum_{b=1}^{B}\text{pbinom}(y, n-r_n, \widehat p^{\ast\ast}_b)$,
respectively, and that direct- and GPQ-bootstrap prediction bounds correspond
to quantiles from these predictive distributions.
Consequently, because $\widehat{p}_{b}^{*}$ and $\widehat{p}_b^{**}$ are small (e.g., less than 0.25) while $\widehat p^\ast_b$ is generally larger than $\widehat p^{\ast\ast}_b$ in Figure~\ref{fig:2},
then $G^{DB}_{Y_n}(y|\boldsymbol{D}_n)$ is generally smaller than $G^{GPQ}_{Y_n}(y|\boldsymbol{D}_n)$,
implying quantiles from $G^{DB}_{Y_n}(y|\boldsymbol{D}_n)$ can be expected to exceed
those from $G^{GPQ}_{Y_n}(y|\boldsymbol{D}_n)$ in data cases with a limited number of events.
However, asymptotically, both $\widehat p_n^\ast$ and $\widehat p_n^{\ast\ast}$ are similarly
normally distributed and symmetric around $\widehat p_n$
 (shown in online supplementary material),
 so that the direct- and GPQ-bootstrap prediction bounds may be expected to behave alike in data situations with a larger number of events
 and larger sample sizes, as seen in Figure~\ref{threeMethods} (and in the Product A application).

\section{Choice of a Distribution}
\label{choice-of-dist}
Extrapolation is usually required when predicting the number of future events based on an on-going time-to-event process. For example, it may be necessary to predict the number of returns in a three-year warranty period based on field data for the first year of operation of a product. An exception arises when life can be modeled in terms of use (as opposed to time in service) and there is much variability in use rates among units in the population. The high-use units will fail early and provide good information about the upper tail of the amount-of-use return-time distribution (e.g., \citet{hong2010}). 

When extrapolation is required, predictions can be strongly dependent on the distribution choice.
In most applications, especially with heavy censoring, there is little or no useful information in the data to help choose a distribution.
Then, for example, it is best to choose a failure-time distribution based on knowledge of the failure mechanism and the related physics/chemistry of failure. In important applications, this would be typically be done by consulting with experts who have such knowledge.

For example, the lognormal distribution could be justified for failure times that arise from the product of a large number of small, approximately independent positive random quantities.  Examples include failure from crack initiation and growth due to cyclic stressing of metal components (e.g., in aircraft engines) and chemical degradation like corrosion (e.g., in microelectronics). These are two common applications where the lognormal distribution is often used. \citet[][pages 36-37]{GnedenkoBelyayevSolovyev1969} provide mathematical justification for this physical/chemical motivation.

\begin{figure}[t!]
	\begin{tabular}{cc}
		\includegraphics[width=0.5\linewidth]{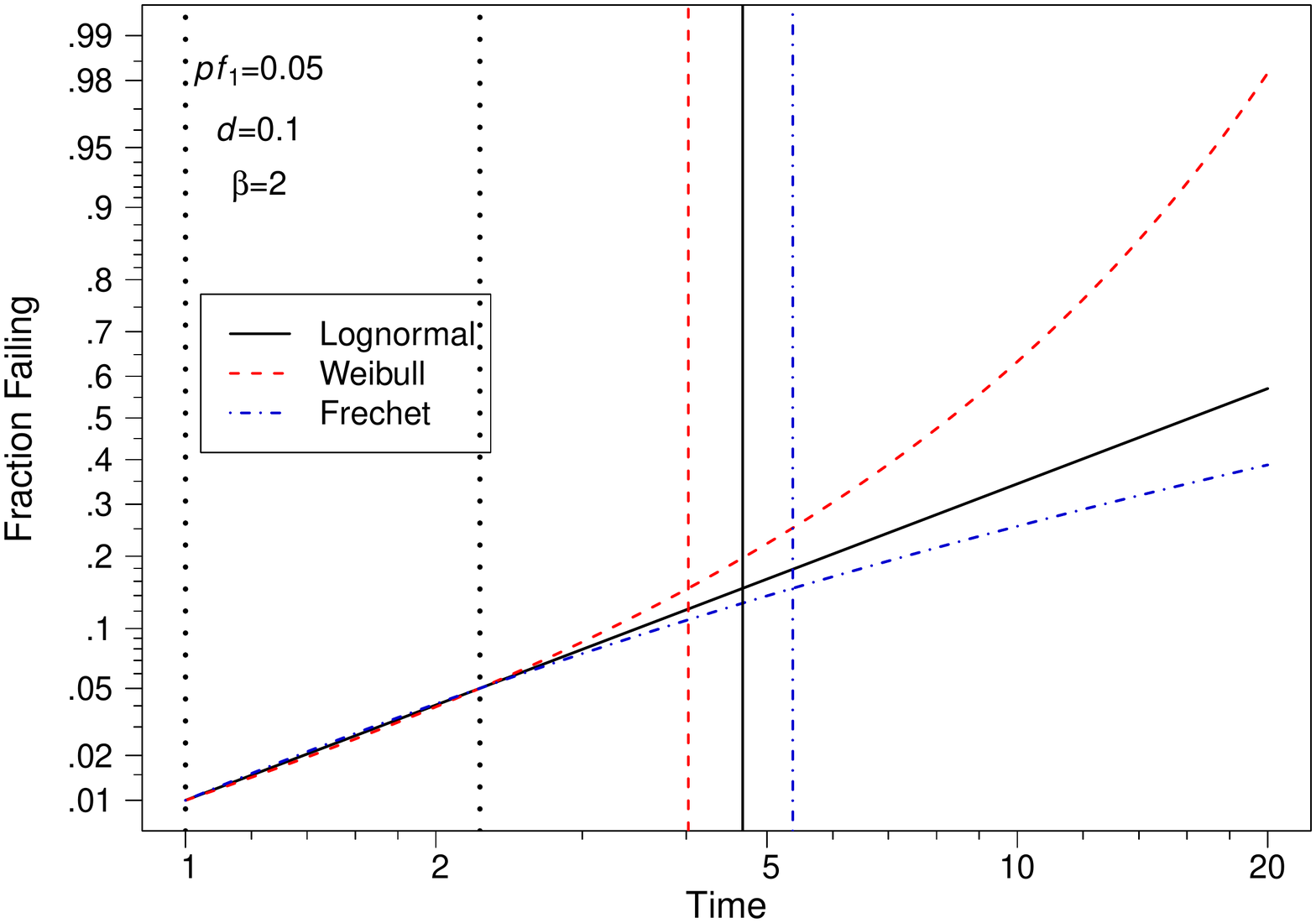} & \includegraphics[width=0.5\linewidth]{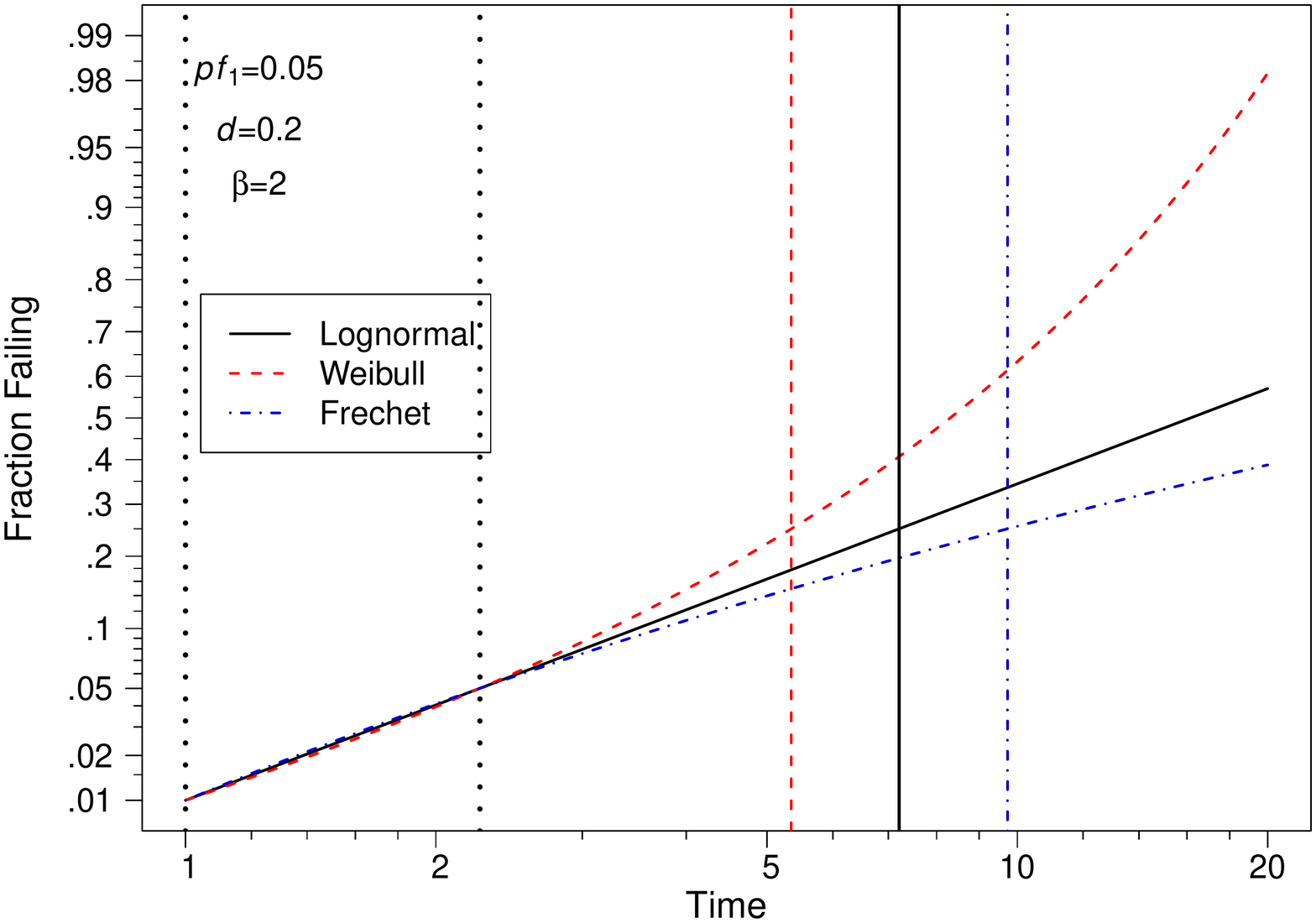} \\
		\includegraphics[width=0.5\linewidth]{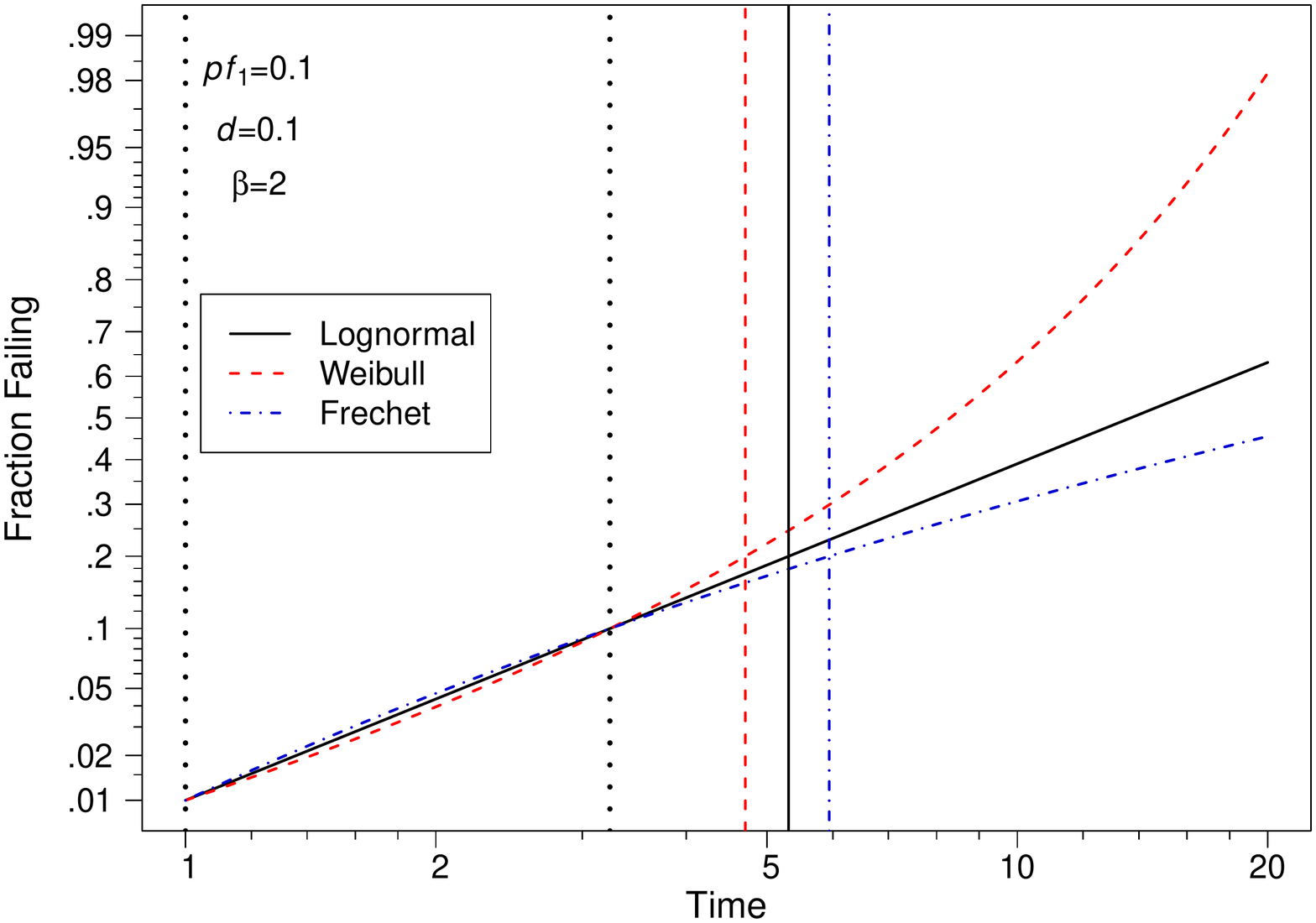} &
		\includegraphics[width=0.5\linewidth]{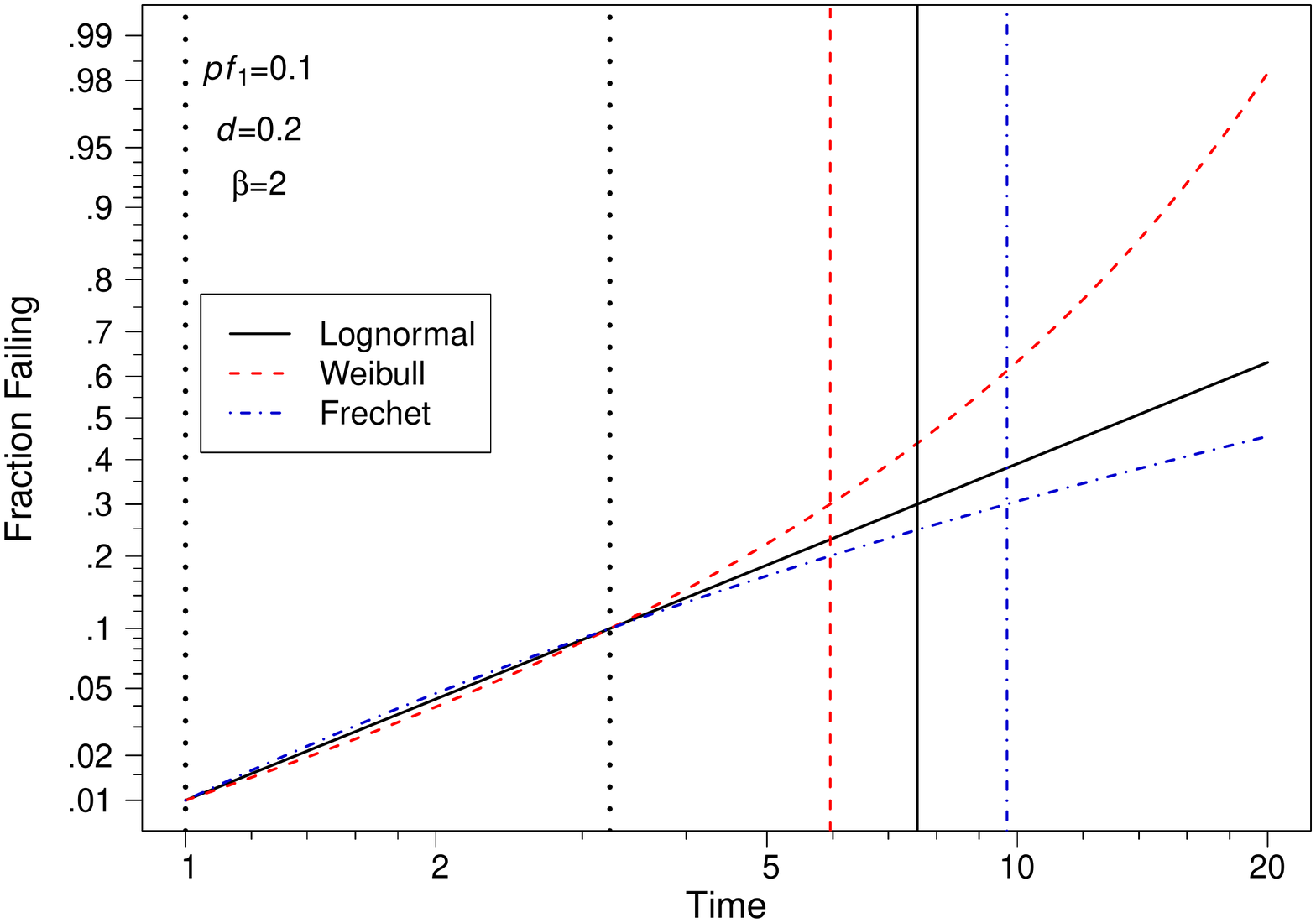} \\
		\includegraphics[width=0.5\linewidth]{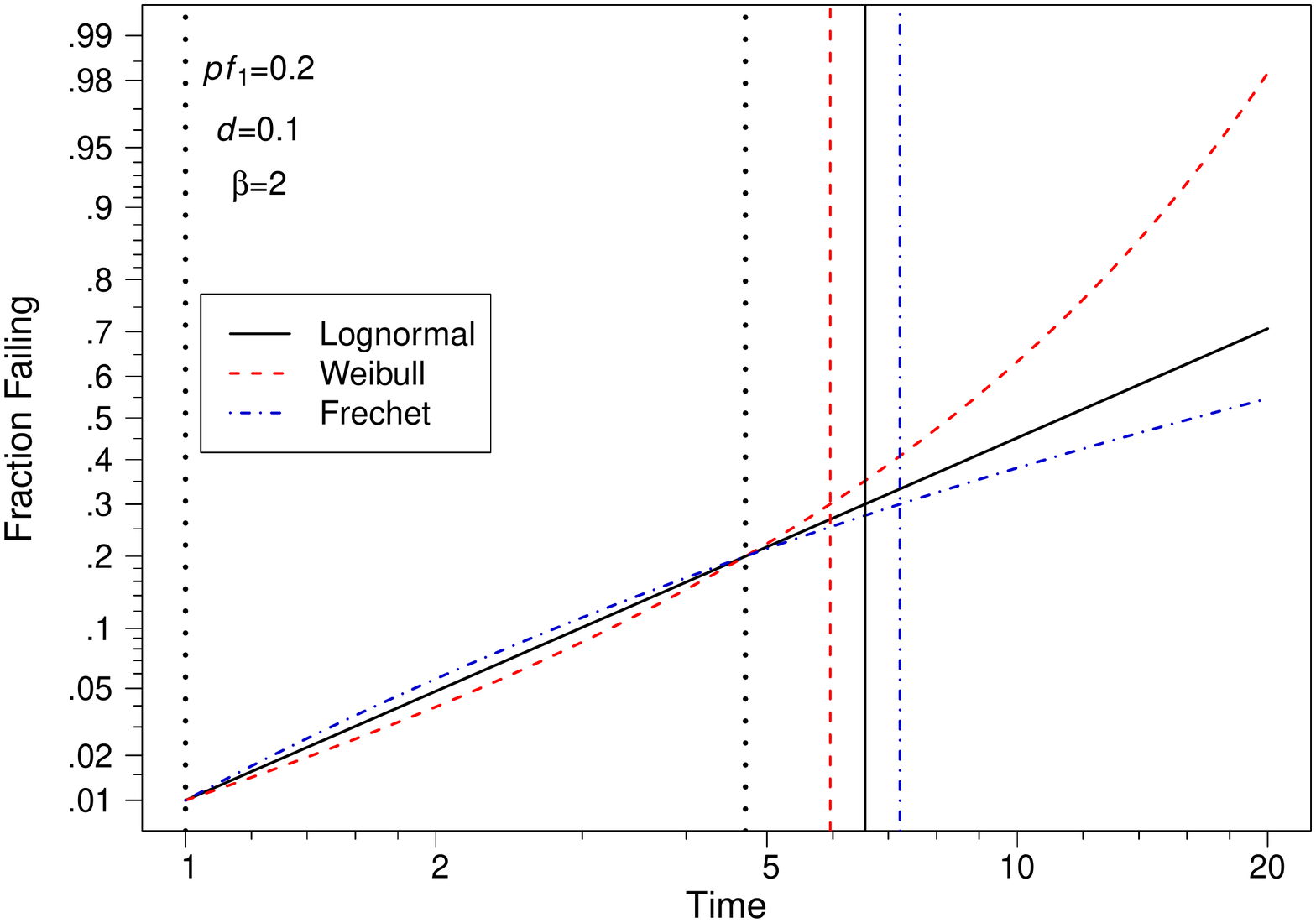} &
		\includegraphics[width=0.5\linewidth]{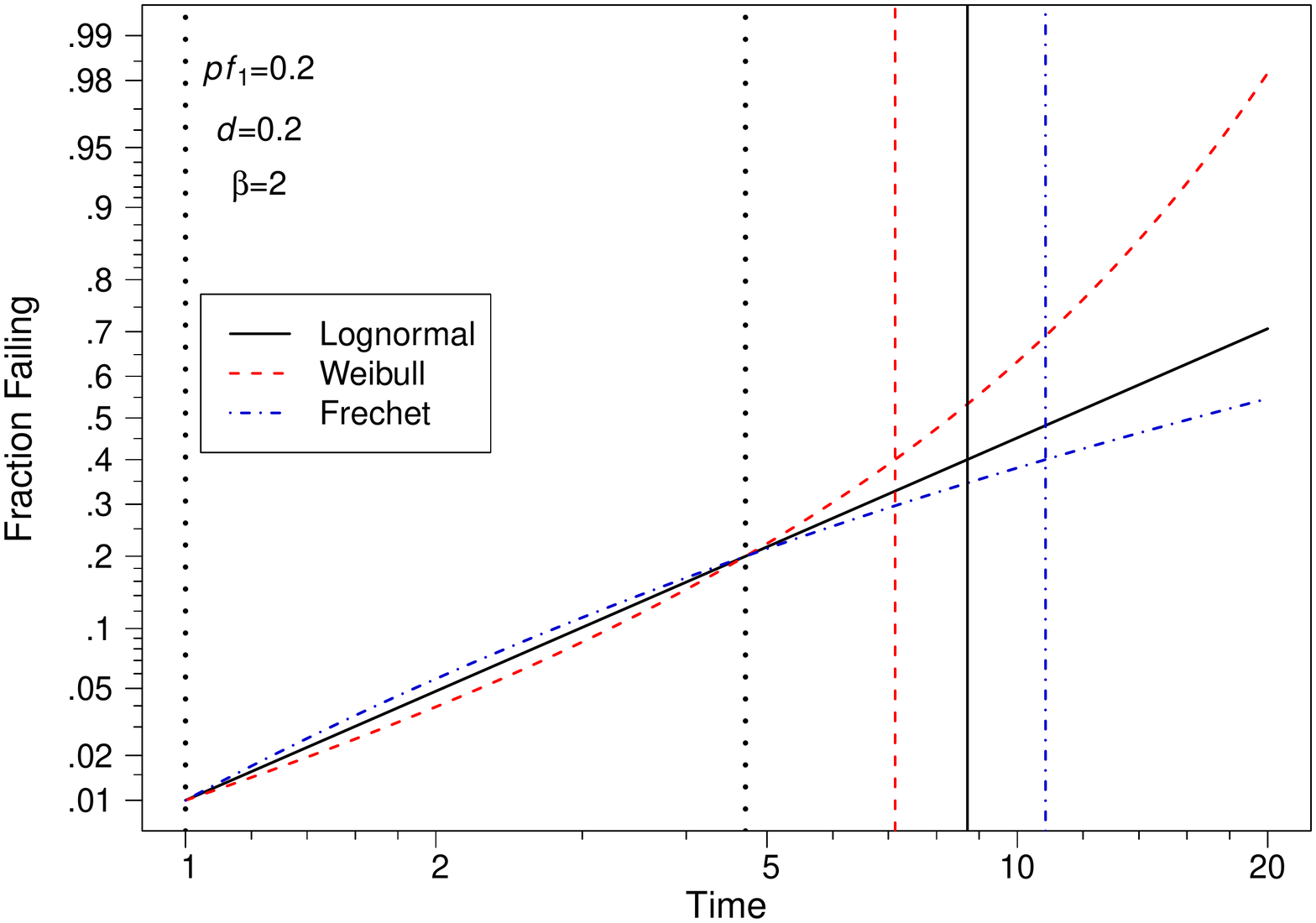} 
	\end{tabular}
	\caption{Distributional comparisons for $\beta=2$. The two vertical
		dotted lines on the left indicate the points in time where all
		three distributions have the same $0.01$ and $p_{f1}$ quantiles.
		The three vertical  lines on the right indicate the times at
		$p_{f2}=p_{f1}+d$ for the three distributions. }
	\label{figure:beta.two}
\end{figure}

Based on extreme value theory, the Weibull distribution can be used to model the distribution of the minimum of a large number of approximately iid positive random variables from certain classes of distributions.  For example, the Weibull distribution may provide a suitable model for the time to first failure of a large number of similar components in a system. Consider a chain with many nominally identical links and suppose that the chain is subjected cyclic stresses over time. As suggested in the previous paragraph, the number of cycles to failure for each link could be described adequately with a lognormal distribution. The chain, however, fails when the first link fails. The limiting distribution of (properly standardized) minima of iid lognormal random variables is a type 1 smallest extreme value (or Gumbel) distribution. For all practical purposes, however, the Weibull distribution provides a better approximation. For further information on this result from the penultimate theory of extreme values, see \citet{Green1976}, \citet[Section 3.11]{Castillo1988}, and \citet{GomesHaan1999}.
Similarly, if failures are driven by the maximum of a large number of approximately iid positive random variables, a Fr\'{e}chet distribution would be suggested. The reciprocal of a Weibull random variable has a Fr\'{e}chet distribution.

Of course, choosing a distribution based on failure-mechanism knowledge is not always
possible. The alternative is to do sensitivity analyses, using different distributions.
Figure~\ref{figure:beta.two} provides a comparison of the Weibull, lognormal, and Fr\'{e}chet
cdfs where the Weibull distribution was chosen with a shape parameter $\beta=2$ and the other factor level combinations of   $p_{f1}$ and $d$ used in the
Section~\ref{simu:study} simulation.
The scale parameter $\eta$ is determined by letting the 0.01 Weibull quantile be 1.
The cdfs are plotted on lognormal probability scales
where the lognormal cdf is a straight line.  The particular parameters for the lognormal
and Fr\'{e}chet distributions were chosen such that the distributions cross at the 0.01 and $p_{f1}$ quantiles, simulating the range of the data where the agreement among
distributions will be good. Similar plots for $\beta=1$ and $\beta=4$ are provided in the
online supplementary material. The Weibull distribution is always more pessimistic
(conservative) than the lognormal and the Fr\'{e}chet is always more optimistic than the
lognormal.
For example, if the true distribution is Weibull but lognormal distribution is used to
fit the data, the prediction intervals, regardless of the method, will underpredict the number of events.
When in doubt, the Weibull distribution is often used because it is the
conservative choice.

\section{Concluding Remarks}
\label{sec:conclusion}
This paper studies the problem of predicting the future
number of events based on censored time-to-event data (e.g., failure times).
This type of prediction is known as within-sample prediction.
A regular prediction problem is defined for which standard plug-in
estimation commonly applies, and it is shown that the within-sample
prediction is not regular and that the plug-in method fails to
produce asymptotically valid prediction bounds.
The irregularity of within-sample prediction and the failure of
the plug-in method motivated the study of the calibration method
as an alternative approach for prediction bounds, though the previously established
theory for calibration bounds does not apply to within-sample prediction.
The calibration method is implemented via bootstrap and called calibration-bootstrap method,
which is proved to be asymptotically correct (i.e., producing prediction
bounds with asymptotically correct coverage).
Then, turning to formulations of a predictive distribution, we study and
validate two other methods to obtain prediction bounds,
namely the direct-bootstrap and GPQ-bootstrap methods.
All prediction methods considered can be applied to both single-cohort
and multiple-cohort data.

While theoretical results show that the calibration-bootstrap method and the
two predictive-distribution-based methods are all asymptotically correct, the
simulation study shows that the direct-bootstrap and GPQ-bootstrap methods
outperform the calibration-bootstrap method in terms of coverage probability accuracy
relative to a nominal coverage level. The two predictive-distribution-based
methods are also easier to implement compared to the
calibration-bootstrap method, and can also be computationally more stable
(e.g., heat exchanger data example). Thus, we recommend predictive
distribution methods, especially the direct-bootstrap
method for general applications involving within-sample prediction.

In this paper, all of the units in the population were assumed to
have the same time-to-event distributions. In many applications,
however, units are exposed to different operating or environmental
conditions, resulting in different time-to-event distributions.
For example, during 1996-2000, the Firestone tires installed on
Ford Explorer SUVs experienced unusually high rates of failure,
where problems first arose in Saudi Arabia, Qatar, and Kuwait because
of the high temperatures in those countries
(see \citet{national2001engineering}).
Having prediction intervals that use covariate information
(like temperature and moisture) could be useful for manufacturers
and regulators in making decisions about a possible product recall,
for example. Similarly, there can be seasonality effects in
time-to-event processes and within-sample predictions.

The methods described in this paper can be extended to handle
either constant covariates or time-varying covariates. Using
calibration-bootstrap methods, \citet{hong2009} used constant
covariates to predict power-transformer failures. Despite
the complicated nature of their data (random right censoring
and truncation and combinations of categorical covariates with
small counts in some cells), \citet{hong2009} were able to use
the fractional random-weight method \citep[e.g.,][]{XuGotwaltHongKingMeeker2020}
to generate bootstrap estimates. \citet{ShanHongMeeker2020} used
time-varying covariates to account for seasonality in two different
warranty prediction applications. As mentioned by one of the
referees, if there is seasonality and data from only part of one
year is available, there is a difficulty. In such cases, it
would be necessary to use past data on a similar process to
provide information about the seasonality.

Covariate information in reliability field data has not been common,
but that is changing, due to a reduction in costs and advances and
in sensor, communications, and storage technology. In the future,
much more covariate information on various system operating/environmental
variables will be available to make better predictions,
as described in \citet{MeekerHong2014}.

\section*{Acknowledgments}
We would like to thanks Luis A. Escobar for helpful comments on this paper.
We are also grateful to the editorial staff, including two reviewers, for
helpful comments that improved the manuscript.
Research was partially supported by NSF DMS-2015390.


\begingroup
	\setlength{\bibsep}{12pt}
	\linespread{1}\selectfont
	\bibliographystyle{apalike}
	\bibliography{reference}  

\begin{thebibliography}{}

\bibitem[Abernethy et~al., 1983]{weibullhandbook}
Abernethy, R., Breneman, J., Medlin, C., and Reinman, G. (1983).
\newblock {\em Weibull Analysis Handbook}.
\newblock Wright-Patterson AFB, Ohio 45433.
\newblock Available at
  \url{https://apps.dtic.mil/dtic/tr/fulltext/u2/a143100.pdf}. Last accessed on
  March 3, 2020.

\bibitem[Aitchison, 1975]{aitchison1975}
Aitchison, J. (1975).
\newblock {Goodness of prediction fit}.
\newblock {\em Biometrika}, 62:547--554.

\bibitem[Atwood, 1984]{atwood1984}
Atwood, C.~L. (1984).
\newblock Approximate tolerance intervals, based on maximum likelihood
  estimates.
\newblock {\em Journal of the American Statistical Association}, 79:459--465.

\bibitem[Barndorff-Nielsen and Cox, 1996]{barncox1996}
Barndorff-Nielsen, O.~E. and Cox, D.~R. (1996).
\newblock Prediction and asymptotics.
\newblock {\em Bernoulli}, 2:319--340.

\bibitem[Beran, 1990]{beran1990}
Beran, R. (1990).
\newblock Calibrating prediction regions.
\newblock {\em Journal of the American Statistical Association}, 85:715--723.

\bibitem[Castillo, 1988]{Castillo1988}
Castillo, E. (1988).
\newblock {\em Extreme Value Theory in Engineering (Statistical Modeling and
  Decision Science)}.
\newblock Academic Press.

\bibitem[Cox, 1975]{cox1975}
Cox, D.~R. (1975).
\newblock {Prediction intervals and empirical Bayes confidence intervals}.
\newblock {\em Journal of Applied Probability}, 12:47–55.

\bibitem[Davison, 1986]{davison1986}
Davison, A.~C. (1986).
\newblock {Approximate predictive likelihood}.
\newblock {\em Biometrika}, 73:323--332.

\bibitem[Escobar and Meeker, 1999]{elawqm1999}
Escobar, L.~A. and Meeker, W.~Q. (1999).
\newblock Statistical prediction based on censored life data.
\newblock {\em Technometrics}, 41:113--124.

\bibitem[Fonseca et~al., 2012]{fonseca2012}
Fonseca, G., Giummolè, F., and Vidoni, P. (2012).
\newblock Calibrating predictive distributions.
\newblock {\em Journal of Statistical Computation and Simulation},
  84:373–383.

\bibitem[Gnedenko et~al., 1969]{GnedenkoBelyayevSolovyev1969}
Gnedenko, B.~V., Belyayev, Y.~K., and Solovyev, A.~D. (1969).
\newblock {\em Mathematical methods of reliability theory}.
\newblock Academic Press.

\bibitem[Gomes and de~Haan, 1999]{GomesHaan1999}
Gomes, M.~I. and de~Haan, L. (1999).
\newblock Approximation by penultimate extreme value distributions.
\newblock {\em Extremes}, 2:71--85.

\bibitem[Green, 1976]{Green1976}
Green, R.~F. (1976).
\newblock Partial attraction of maxima.
\newblock {\em Journal of Applied Probability}, 13:159--163.

\bibitem[Hall et~al., 1999]{hall1999}
Hall, P., Peng, L., and Tajvidi, N. (1999).
\newblock On prediction intervals based on predictive likelihood or bootstrap
  methods.
\newblock {\em Biometrika}, 86:871--880.

\bibitem[Hannig et~al., 2006]{hannig2006}
Hannig, J., Iyer, H., and Patterson, P. (2006).
\newblock Fiducial generalized confidence intervals.
\newblock {\em Journal of the American Statistical Association}, 101:254--269.

\bibitem[Harris, 1989]{harris1989}
Harris, I.~R. (1989).
\newblock Predictive fit for natural exponential families.
\newblock {\em Biometrika}, 76:675--684.

\bibitem[Hong, 2013]{hongpoisson2013}
Hong, Y. (2013).
\newblock {On computing the distribution function for the Poisson-binomial
  distribution}.
\newblock {\em Computational Statistics and Data Analysis}, 59:41–51.

\bibitem[Hong and Meeker, 2010]{hong2010}
Hong, Y. and Meeker, W.~Q. (2010).
\newblock Field-failure and warranty prediction based on auxiliary use-rate
  information.
\newblock {\em Technometrics}, 52:148--159.

\bibitem[Hong and Meeker, 2013]{hong2013}
Hong, Y. and Meeker, W.~Q. (2013).
\newblock Field-failure predictions based on failure-time data with dynamic
  covariate information.
\newblock {\em Technometrics}, 55:135–149.

\bibitem[Hong et~al., 2009]{hong2009}
Hong, Y., Meeker, W.~Q., and McCalley, J.~D. (2009).
\newblock Prediction of remaining life of power transformers based on left
  truncated and right censored lifetime data.
\newblock {\em The Annals of Applied Statistics}, 3:857--879.

\bibitem[Lawless and Fredette, 2005]{lawless2005}
Lawless, J.~F. and Fredette, M. (2005).
\newblock Frequentist prediction intervals and predictive distributions.
\newblock {\em Biometrika}, 92:529--542.

\bibitem[Meeker and Hong, 2014]{MeekerHong2014}
Meeker, W.~Q. and Hong, Y. (2014).
\newblock Reliability meets big data: opportunities and challenges.
\newblock {\em Quality Engineering}, 26:102--116.

\bibitem[{National Highway Traffic Safety Administration},
  2001]{national2001engineering}
{National Highway Traffic Safety Administration} (2001).
\newblock Engineering analysis report and initial decision regarding
  {EA00-023}: {Firestone Wilderness AT Tires}.
\newblock {\em Washington, DC: US Department of Transportation}.
\newblock Available at
  \url{https://icsw.nhtsa.gov/nhtsa/announce/press/Firestone/firestonesummary.html}.
  Last accessed on March 3, 2020.

\bibitem[Nelson, 2000]{nelson2000}
Nelson, W. (2000).
\newblock Weibull prediction of a future number of failures.
\newblock {\em Quality and Reliability Engineering International}, 16:23--26.

\bibitem[Scholz, 2001]{scholz1996maximum}
Scholz, F. (2001).
\newblock {Maximum likelihood estimation for Type I censored Weibull data
  including covariates}.
\newblock In {\em ISSTECH-96-022, Boeing Information \& Support Services, PO
  Box 24346, MS-7L-22}.
\newblock Available at
  \url{http://faculty.washington.edu/fscholz/DATAFILES498B2008/ISSTECH-96-022.pdf}.
  Last accessed on August 3, 2020.

\bibitem[Shan et~al., 2020]{ShanHongMeeker2020}
Shan, Q., Hong, Y., and Meeker, W.~Q. (2020).
\newblock Seasonal warranty prediction based on recurrent event data.
\newblock {\em Annals of Applied Statistics}, 14:929--955.

\bibitem[Shen et~al., 2018]{shen_liu_xie_2018}
Shen, J., Liu, R.~Y., and Xie, M.-G. (2018).
\newblock Prediction with confidence—a general framework for predictive
  inference.
\newblock {\em Journal of Statistical Planning and Inference}, 195:126–140.

\bibitem[Wang et~al., 2012]{wang2012}
Wang, C., Hannig, J., and Iyer, H.~K. (2012).
\newblock Fiducial prediction intervals.
\newblock {\em Journal of Statistical Planning and Inference}, 142:1980–1990.

\bibitem[Xie and Singh, 2013]{xie_singh_2013}
Xie, M.-G. and Singh, K. (2013).
\newblock Confidence distribution, the frequentist distribution estimator of a
  parameter: A review.
\newblock {\em International Statistical Review}, 81:3–39.

\bibitem[Xu et~al., 2020]{XuGotwaltHongKingMeeker2020}
Xu, L., Gotwalt, C., Hong, Y., King, C.~B., and Meeker, W.~Q. (2020).
\newblock Applications of the fractional-random-weight bootstrap.
\newblock {\em The American Statistician}.
\newblock \url{https://doi.org/10.1080/00031305.2020.1731599}.

\end{thebibliography}
\endgroup
\end{document}